\documentclass[aps,prd,floatfix,nofootinbib,superscriptaddress,preprintnumbers,amsmath,amssymb,onecolumn,notitlepage]{revtex4-1}
\usepackage{graphicx}
\usepackage{textcomp}
\usepackage{mathtools}
\usepackage{amsmath}
\usepackage{amssymb}
\usepackage{esint}
\usepackage{comment}
\usepackage[utf8]{inputenc}
\usepackage{hyperref}
\usepackage{subcaption}
\usepackage{color}
\usepackage{MnSymbol}

\newcommand{\myref}{Ref.~}

\newcommand{\as}{\alpha_s}
\newcommand{\Qs}{Q_s}

\renewcommand{\v}[1]{ \ensuremath{ {\mathbf{#1}_\perp} }}
\renewcommand{\vec}[1]{ \ensuremath{ {\mathbf{#1}} }}
\renewcommand{\nabla}{ \ensuremath{\partial}}

\begin{document}

\title{Initial state description of azimuthally collimated long range correlations in ultrarelativistic light-heavy ion collisions}
\author{Mark Mace}
\affiliation{Department of Physics,  P.O. Box 35, 40014, University of Jyv\"{a}skyl\"{a}, Finland}
\affiliation{Helsinki Institute of Physics, P.O. Box 64, 00014 University of Helsinki, Finland}
\affiliation{Physics Department, Brookhaven National Laboratory, Upton, New York 11973-5000, USA}
\affiliation{Department of Physics and Astronomy, Stony Brook University, Stony Brook, NY 11794, USA}
\author{Vladimir V. Skokov}
\affiliation{Department of Physics, North Carolina State University, Raleigh, North Carolina 27695, USA}
\affiliation{RIKEN-BNL Research Center, Brookhaven National Laboratory, Upton, New York 11973-5000, USA}
\author{Prithwish Tribedy}
\affiliation{Physics Department, Brookhaven National Laboratory, Upton, New York 11973-5000, USA}
\author{Raju Venugopalan}
\affiliation{Physics Department, Brookhaven National Laboratory, Upton, New York 11973-5000, USA}

\date{\today}
\begin{abstract}
It was argued in arXiv:1805.09342 and arXiv:1807.00825 that the systematics of the azimuthal anisotropy coefficients $v_{2,3}$ measured in ultrarelativistic light-heavy ion collisions at RHIC and the LHC can be described in an initial state dilute-dense Color Glass Condensate (CGC) framework. We elaborate here on the discussion in these papers and provide further novel results that strengthen their conclusions. The underlying mathematical framework and numerical techniques employed are very similar to those in the CGC based IP-Glasma model used previously as initial conditions for heavy-ion collisions. The uncertainties in theory/data comparisons for small systems are discussed, with unknowns that are specific to the model distinguished from those that are generic to all models. We present analytical arguments that demonstrate that quantum interference effects such as Bose enhancement and Hanbury-Brown-Twiss correlations of gluons, as well as coherent multiple scattering of gluons in the projectile off color domains in the target, are enhanced in rare events.  The quantum origins of the large anisotropies in small systems are corroborated by numerical results for deuteron-gold collisions that show that large anisotropies in rare configurations can occur when the nucleons in the projectile overlap significantly. This is at variance with the classical intuition of hydrodynamical models. We also comment on the consequences of ignoring the many-body color charge correlations of gluons in models that only consider geometrical fluctuations in the energy density. 
\end{abstract}

\maketitle

\section{Introduction} 

In two recent papers~\cite{Mace:2018vwq,Mace:2018yvl}, we presented novel results, based on the Color Glass Condensate (CGC) Effective Field
Theory (EFT)~\cite{Gelis:2010nm}, for the coefficients $v_{2,3}$ of azimuthally anisotropic long range in rapidity two-particle correlations measured in light ion-heavy ion collisions at RHIC and LHC respectively. In the first of these papers Ref.~\cite{Mace:2018vwq}, we compared our results to data presented by the PHENIX collaboration at RHIC in collisions of a proton/deuterium/helium-3 nucleus with a gold nucleus ($p$+Au/$d$+Au/$^3$He+Au) at center-of-mass energy of $\sqrt{s}=200$ GeV/nucleon~\cite{PHENIX:2018lia,Aidala:2018mcw}, and showed that semi-quantitative agreement with the systematics of the data was achieved. In Ref.~\cite{Mace:2018yvl}, we considered $\sqrt{s}=5.02$ TeV/nucleon proton-lead collisions at the LHC, and demonstrated how a simple power counting argument describes the systematics of the correlations as a function of the charged particle multiplicity. These power counting arguments were also fortified by numerical computations in Ref.~\cite{Mace:2018yvl}. 

The computations in Refs.~\cite{Mace:2018vwq,Mace:2018yvl} were performed within the dilute-dense framework of the CGC~\cite{Kovchegov:1998bi,Dumitru:2001ux,Blaizot:2004wu}; in this framework,  gauge fields produced in heavy-ion collisions are obtained from solutions of the QCD Yang-Mills equations to lowest and first subleading  contributing order in the color source density of the projectile and to all orders in that of the target. This dilute-dense framework can be obtained as a systematic expansion in the color charge densities in the projectile and target nucleus of the more general dense-dense CGC formalism, wherein all orders are retained in the color charge densities of both projectile and target. The latter is appropriate for the collisions of heavy-nuclei\footnote{The power counting in the CGC is in the ratios $ g^2 \rho_{\rm pr,t}/k_{\perp {\rm pr,t}}^2$, where $\rho_{\rm pr(t)}$ corresponds to the color charge density of the projectile (target) and $k_{\perp {\rm pr(t)}}$ denotes the transverse momenta transferred from the projectile (target) to the production of an inclusive final state. Here $g$ is the QCD coupling.}. The Yang-Mills equations in the dense-dense framework can only be solved numerically, and the solutions, corresponding to the real-time evolution of Glasma gauge fields, are obtained using classical-statistical methods in lattice gauge theory~\cite{Krasnitz:1998ns,Krasnitz:1999wc,Krasnitz:2001qu,Krasnitz:2003jw,Lappi:2003bi}. Double inclusive correlations in the dense-dense framework were first computed in \cite{Lappi:2009xa}. This 2+1-D Glasma framework is however incomplete as it neglects the full 3+1D Yang-Mills dynamics; this 3+1-D dynamics has a qualitative impact on thermalization and has been shown~\cite{Berges:2013eia,Berges:2013fga,Berges:2014bba} to lead to the late time kinetic framework of ``bottom up" thermalization~\cite{Baier:2000sb}.

The IP-Glasma model~\cite{Schenke:2012wb,Schenke:2012fw} provides a numerical implementation of the transverse dynamics of color charge densities in this dense-dense framework;  it includes phenomenological constraints on the color charge distributions in the projectile and target. In particular, the color charge distributions in a proton are extracted from saturation model fits to HERA deeply inelastic scattering (DIS) data~\cite{Rezaeian:2012ji}.  To generate nuclear color charge distributions, nucleons are sampled from the projectile and target wavefunctions using standard Monte Carlo techniques. This procedure generates event-by-event color charge distributions in both projectile and target which play the role of color sources in the Yang-Mills equations. The solutions of these equations generate the energy densities and collective four-velocities, which in turn are employed as initial conditions for hydrodynamic evolution, as described by the MUSIC model~\cite{Schenke:2010rr}. The resulting 2+1-D IP-Glasma+MUSIC model provides an excellent phenomenological description of data in heavy-ion collisions~\cite{Gale:2012rq} albeit, as noted, it suffers from systematic uncertainties in the treatment of preequilibrium dynamics. 

As we shall discuss further, the numerical procedure we follow in setting up our problem is identical to that of the IP-Glasma model in nearly all aspects. The primary difference is that in the dilute-dense approximation one obtains analytical expressions for the Yang-Mills gluon fields after the collision in terms of the color charge densities in the projectile and target. This provides a tremendous simplification in computational difficulty which was exploited in Refs.~\cite{Mace:2018vwq,Mace:2018yvl} to compute the systematics of multiparticle correlations. 

Even though analytical expressions for the gauge fields  are available in the dilute-dense framework, their dependence on the color charge densities in the projectile and target is highly nonlinear. It is therefore difficult to intuit in general from the corresponding expressions for azimuthal correlations their nontrivial behavior as a function of the event-by-event locations of the projectile and target color charge distributions. The intuition gained from simpler models is therefore valuable even if not necessarily the full picture. 

As an example, the possibility that initial state momentum correlations could reproduce the systematics of the $v_n$ n-th harmonic coefficients of two particle~\cite{Lappi:2015vha,Lappi:2015vta} and four particle~\cite{Dusling:2017dqg,Dusling:2017aot,Dusling:2018hsg,Fukushima:2017mko,Zhang:2019yhk} correlations in small systems was previously discussed in a proof-of-principle parton model  of independent eikonal partons scattering off color domains in a nuclear target~\cite{Kovner:2010xk,Dumitru:2014yza}. In this simple model, the nontrivial azimuthal momentum correlations generated between the partons depends on the coherent chromo electric and magnetic fields localized on distance scales of the order of the inverse saturation scale $1/\Qs$ in the target. For an overlap area $S_\perp$ of the projectile with the target, one might anticipate a $1/\Qs^2 S_\perp$ inverse domain scaling. However Refs.~\cite{Dusling:2017dqg,Dusling:2017aot,Dusling:2018hsg} demonstrated that this scaling only occurs for large transverse momentum kicks $p_\perp > \Qs$ from the target color-electric fields to the incoming collinear (on-shell) partons. For $p_\perp < \Qs$, the partons interact coherently with color fields spanning several domains resulting in  azimuthal correlations that are only weakly dependent of $\Qs^2 S_\perp$. A more sophisticated variant of this model was employed recently to describe the PHENIX $v_{2,3}$ data on $p$+A collisions~\cite{Davy:2018hsl}.

Although the parton model described in \cite{Lappi:2015vha,Lappi:2015vta,Dusling:2017dqg,Dusling:2017aot,Dusling:2018hsg} provides an intuitive picture of azimuthal correlations induced by color domains in the target, it misses some features of the full range of correlations that are present in the CGC framework of high energy scattering. An important ingredient that was missing in this simple  model was the quantum statistics of the incoming and outgoing  partons, which are predominantly gluons. These are responsible to additional quantum interference contributions arising from both the Bose enhancement of gluons in the hadron wavefunctions as well as Hanbury-Brown--Twiss (HBT) correlations between the produced gluons~\cite{Dumitru:2008wn,Kovchegov:2012nd,Altinoluk:2015uaa,Altinoluk:2015eka}. These effects, in addition to the coherent scattering off color domains, are contained in the dilute-dense CGC framework and contribute to the azimuthal correlations seen in our numerical results presented in Refs.~\cite{Mace:2018vwq,Mace:2018yvl}. We will provide a qualitative discussion of these effects and shall also discuss numerical results that provide further insight into their quantum origins. 

The paper is organized as follows. In the next section, we will provide a theoretical summary of the dilute-dense formalism for inclusive multiparticle production from the CGC with emphasis on 
how one computes the azimuthal anisotropy coefficients  $v_{n}$. Extensions of this framework are also discussed briefly. The implementation of the formalism is nontrivial and is discussed in detail in Section 3. We provide the recipe to compute the event-by-event color charge distributions. Even though analytical expressions for the gluon fields are known, numerical results for these given color charge distributions require lattice gauge theory techniques. These are identical to their implementation in earlier work~\cite{Krasnitz:2003jw,Lappi:2003bi,Lappi:2007ku,Schenke:2012fw} but are spelled out here for the problem of interest. We end this section with a discussion of the particulars of our implementation of multiparticle production. 

In Section 4, we present results from our numerical computations. We first present results for the n-particle multiplicity distribution in both proton-proton ($p$+$p$) and light-heavy ion collisions. As argued previously~\cite{Gelis:2009wh} , color charge fluctuations in the McLerran-Venugopalan (MV) model~\cite{McLerran:1993ka,McLerran:1993ni,McLerran:1994vd} generate negative binomial distributions; this is a natural consequence of the Bose statistics of gluons. In general, in the CGC EFT, there can be additional sources of fluctuations in the saturation scale $Q_s$~\cite{Iancu:2007st,Dumitru:2017cwt,Kovner:2018azs}. For light-heavy ion collisions, there are additional fluctuations due to fluctuations in nucleon positions. We show that the multiplicity distributions in both proton-proton and deuteron-gold collisions can be reproduced in our framework. The model fits to multiplicity distributions fix nonperturbative parameters in the model, though not uniquely. We next compare the measured $p_\perp$ dependence of charged pions in high multiplicity events to the computed single inclusive gluon distribution. At low $p_\perp$, the shape of the two distributions is similar if one assumes a constant gluon to hadron conversion factor; however at $p_\perp \geq 1$ GeV the gluon distribution rapidly overshoots the data. We stress that the comparison of the two distributions requires strong assumptions regarding fragmentation in all approaches. 

We next show results for $v_{2,3}$ as a function of the transverse momentum $p_\perp$ for the $0-5$\% centrality class in $p$+Au, $d$+Au and $^3$He+Au collisions at the highest RHIC center of mass energy. These results were presented previously in Ref.~\cite{Mace:2018vwq}; here we present our results with error bands based on known sources of theoretical uncertainty. 
In \cite{Mace:2018vwq}, we had  conjectured that the results for $v_2(p_\perp)$ for events in the different light-heavy ion systems with the same value of large charged hadron multiplicity $N_{\rm ch}$ would be identical. We will show a prediction for same $N_{\rm ch}$ (albeit different centrality class) events in $d$+Au and $^3$He+Au collisions that is consistent with the conjecture. For comparisons of the same $N_{\rm ch}$ (but lower $N_{\rm ch}$ relative to the $d$+Au versus $^3$He+A comparison) in $p$+Au and $d$+Au, the central numerical values of $v_2(p_\perp)$ differ; however, the results for the two systems overlap when the estimated theory uncertainty is taken into account. These results are consistent with the comparison in the recent PHENIX paper~\cite{PHENIX:2018lia,Aidala:2018mcw}.

In Section 5, we examine further the origins of the azimuthal anisotropies and many-body correlations in small systems. For simplicity, we will consider here only $d$+Au collisions. 
To gain a deeper understanding of the underlying many-body dynamics in the collision, we consider the simpler MV model on the target nucleus side thereby suppressing nucleon fluctuations in the gold nucleus. We present analytical arguments in the MV model that show simply that the enhancement of multiparticle cumulants in 
rare configurations is a nontrivlal consequence of the Bose statistics of gluons. These arguments also indicate that high multiplicity (rare) events should favor deuteron configurations where the proton and the nucleus are closer than their separation in minimum bias (typical) events. These arguments are corroborated by numerical simulations in the MV model. 
We find strikingly that the multiparticle probability distribution as a function of multiplicity in the MV model is qualitatively different from that of the Monte Carlo Glauber (MC Glauber) model~\cite{Miller:2007ri}; in the latter, both rare and typical events are dominated by configurations where the nucleons in the projectile are separated as opposed to overlapping. The same trend as 
the MV model is seen in our full numerical results including all sources of fluctuations; though less dramatic than the MV case, the results are qualitatively different from those in the MC Glauber model. 

Our discussion on multiplicity distributions is very much in line with the discussion in Glauber's quantum theory of optical coherence\footnote{This work has a 
different provenance from that of the Glauber model of multiple scattering~\cite{Franco:1965wi}.} wherein he showed that the quantum statistics of photons generates qualitatively different results for multiparticle cumulants from classical optics because the latter ignores the  ``n-factorial" statistics of photons~\cite{Glauber:1963fi}. Our studies suggest that models that ignore or improperly implement the Bose statistics of multigluon states (and their corresponding color charge fluctuations) miss fundamental features of multiparticle production in QCD. This may not be apparent in 
single inclusive distributions but are essential for the description of many-body correlations. We comment in particular on the consequences for a recent model in the literature that argues that sub-nucleon color charge fluctuations are not pertinent in describing rare events in small systems~\cite{Nagle:2018ybc}.

We next discuss the qualitative trends observed in the $v_2$ azimuthal anisotropy distributions. We first classify 
the $v_2$ obtained in our simulations as a function of $p_\perp$ and the relative separation between the color charge distributions of each of the nucleons in the transverse plane. We observe that the largest contribution to $v_2$ comes from overlapping configurations and that $v_2$ decreases monotonically with increasing transverse separation between the nucleons. This effect is at variance with a simple ``geometrical" hydrodynamic picture where the larger initial spatial anisotropies at larger separations generate larger $v_2$ through flow\footnote{We note however that large shape fluctuations in overlapping configurations may also generate significant $v_2$ if a flow scenario can be justified~\cite{Schenke-Shen-Tribedy}.}.  The monotonic decrease in $v_2$ with nucleon separation is seen for all the $p_\perp$ bins examined but is stronger with increasing $p_\perp$ out to $2$ GeV. This observation is in variance with expectations based on simple domain scaling but appears consistent with the picture of such momentum anisotropies being generated by the overlap of momentum space projectile and target unintegrated gluon distributions~\cite{Dumitru:2010iy,Dusling:2013qoz}. 

In the concluding section, we will summarize our results and discuss open questions and further refinements that are feasible in of our theory framework and the possibility of extracting information on these from experiments.

\section{Theoretical summary} 
\label{sec:theory}

In the light-front formulation of the CGC EFT, a natural kinematic separation exists between the dynamics of fast-moving valence-like partons in the hadron wavefunctions and that of the small x partons, which are primarily gluons. The latter can be treated as classical fields due to their high phase space occupancy within the hadron while the former can be treated as 
static light-cone sources that are highly Lorentz contracted in the direction of their motion~\cite{McLerran:1993ka,McLerran:1993ni,McLerran:1994vd}. The dynamics of the soft gauge fields in the presence of these static light-cone sources is described by the classical QCD Yang-Mills (CYM) equations,
\begin{eqnarray}
[ D_\mu,F^{\mu \nu}]=J^{\nu}\,,
\label{eqn:cym}
\end{eqnarray}
where $J^\mu=\delta^{\mu +} \rho(\mathbf{x}_\perp) \delta(x^-)$ for a hadron with a large momentum $P^+\rightarrow \infty$ 
both before and after the collision. In this framework, the early time dynamics of the collision of two nuclei can be described as the collision of two classical gluon ``shock wave" fields 
which are each sourced by their color charge densities. Since we will be discussing light-heavy ion collisions in this paper, we will henceforth use the abbreviated notations ``${\rm pr}$" and ``${\rm t}$" to refer to the color charge densities in the light and heavy ion respectively. 

Before the collision, for a single configuration of sources, the soft gauge fields created by the hard sources are pure gauges which are discontinuous across the region where the color 
sources have support, and take the form~\cite{McLerran:1993ka,McLerran:1993ni},
\begin{eqnarray}
\partial_i \alpha^i_{{\rm pr},{\rm t}}(\mathbf{x}_\perp)&=& g \rho_{{\rm pr},{\rm t}}(\mathbf{x}_\perp)\,, \\
\alpha^i_{{\rm pr},{\rm t}}(\mathbf{x}_\perp)&=& \frac{-1}{ig} U_{{\rm pr},{\rm t}}(\mathbf{x}_\perp) \partial^i U_{{\rm pr},t}^\dagger(\mathbf{x}_\perp) \label{eqn:LLCalphai_def} \,.
\end{eqnarray}
The solution of the Yang-Mills equation in either the projectile or target before the collision is represented by a path ordered light-like Wilson line 
\begin{equation}
U_{{\rm pr},{\rm t}}(\v{x}) = {\cal P} \exp \left( - i g \int dx^\mp { A}_{\rm pr, t }^{\pm a}(x^\mp, \v{x}) T_a \right)\,
	\label{Eq:U}
\end{equation}
along the light-cones of the respective nuclei, where $A^{\pm a}_{\rm pr}$ is the only nonzero component of the gauge field for the projectile (target) nucleus upon gauge rotation from light-cone gauge to Lorenz gauge,  
$\partial_\perp^2  A^{\pm a}_{\rm pr} =  \rho_{{\rm pr},{\rm t}} ^a(x^\mp,\v{x})$~\cite{JalilianMarian:1996xn,Kovchegov:1996ty,Kovchegov:1997pc}. This Wilson line, through its dependence on the physical sources, embeds a color memory effect which has an exact correspondence to the gravitational memory effect that may in future be measured by gravitational wave detectors~\cite{Ball:2018prg}.

The analogous analytical solutions  $A^\mu$ of the Yang-Mills equations in the forward light-cone {\it after the collision} in terms of the gauge fields $\alpha^i_{\rm pr, t}$ before the collision are not available for the general case where one retains arbitrary powers of the color charge densities in both the projectile and target. Perturbative solutions exist in a dilute-dilute limit where one expands the equations to lowest nontrivial order in $ g^2 \rho_{\rm pr,t}/k_{\perp,{\rm pr,t}}^2 \ll1$, where $k_{\perp,{\rm pr}}$ and $k_{\perp,{\rm t}}$ denote the transverse momentum transfer from the projectile and target respectively in physical processes of interest~\cite{Kovner:1995ja,Gyulassy:1997vt,Kovchegov:1997ke}. (Interestingly, these solutions too bear an exact correspondence to gravitational wave solutions in the ultrarelativistic limit~\cite{Goldberger:2016iau}.)
Similarly, for asymmetric collisions of small projectiles on large targets, or in the kinematic region of projectile forward rapidity for like-sized projectile and target, it may suffice to good accuracy to 
perform computations to the lowest nontrivial order in the projectile color charge density $g^2 \rho_{\rm pr}/k_{\perp,{\rm pr}}^2\ll 1$ while keeping all orders in the target color charge density $g^2 \rho_{\rm t}/k_{\perp,\rm t}^2\sim 1$ in a so-called dilute-dense approximation. As alluded to previously, the full solution of the Yang-Mills equations to all orders in $g^2 \rho_{\rm pr,t}/k_{\perp,{\rm pr,t}}^2$ is only known numerically~\cite{Krasnitz:1998ns,Krasnitz:1999wc,Krasnitz:2001qu,Krasnitz:2003jw,Lappi:2003bi}.

More specifically, the analytical solutions in the dilute-dense approximation are obtained by expanding the Wilson line for the projectile in Eq.~\eqref{eqn:LLCalphai_def} 
 in powers of $\rho_{\rm pr}$~\cite{Kovchegov:1998bi,Dumitru:2001ux,McLerran:2016snu}. In Fock-Schwinger or radiation gauge $A_\tau=x_-  A_+ + x_+ A_-=0$, one obtains
\begin{eqnarray}
\alpha_{\rm pr}^i=g {\partial^i A^{+}_{\rm pr}(\v{x}) }-{\frac{ig^3}{2}\left(\delta_{ij} - \frac{\partial_i \partial_j}{\partial_\perp^2} \right) \left[\partial^j A^{+}_{\rm pr}(\v{x}) ,A^{+}_{\rm pr}(\v{x}) \right]} +\mathcal{O}\left( [A^{+}_{\rm pr}(\v{x})] ^3\right)\,,
\end{eqnarray}
where $A^{+b}_{\rm pr}(\v{x}) = \int dx^- A^{+b}_{\rm pr}(x^-, \v{x}) $. 
The expression for $\alpha_{\rm t}$ is retained in terms of the Wilson line in the target containing all orders in $g^2 \rho_{\rm t}/k_{\perp,\rm t}^2$. The corresponding single inclusive gluon distribution in this expansion in the parameter ${g^2 \rho_{\rm pr}}/{k_\perp^2}\ll 1$, for fixed values of the color charge densities, can be expressed formally as 
\begin{align}
	\frac{dN_g}{d^2 k_\perp dy}\Big[\rho_{\rm pr}, \rho_{\rm t}\Big] = \frac{1}{g^2}  \left[
     \left(g^2 \frac{\rho_{\rm pr}}{k_\perp^2} \right)^2
	  \ f_1 \! \left( g^2  \frac{\rho_{\rm t}}{k_\perp^2} \right) +
    \left( g^2  \frac{\rho_{\rm pr}}{k_\perp^2} \right)^3 \ f_2 \! \left( g^2 \frac{\rho_{\rm t}}{k_\perp^2} \right)  +
    \left( g^2  \frac{\rho_{\rm pr}}{k_\perp^2} \right)^4 \ f_3 \! \left( g^2  \frac{\rho_{\rm t}}{k_\perp^2} \right) + \ldots \right] \,.
  \label{Eq:SIP}
\end{align} 
The function $f_1$ containing all orders $g^2 \frac{\rho_{\rm t}}{k_\perp^2}$ in the color charge density of the target has long been known analytically~\cite{Kovchegov:1998bi,Dumitru:2001ux,Blaizot:2004wu}. The functional forms of $f_i$ for $i\ge 2$ are not known analytically. Even the function $f_1$, as we shall see, 
 involves rather complicated momentum integrals of Wilson lines of the target field. It is therefore necessary to use numerical methods; in particular, we will employ  lattice gauge theory methods which are essential to ensure that the gauge invariance of physical quantities is not broken by numerical artifacts. 
 
 The dilute-dense approximation however is much simpler than the full dense-dense CYM numerical computation because $f_1$ does not require real time evolution of the gauge fields nor does it require numerical implementation of the Lehmann-Symanzik-Zimmermann (LSZ) reduction formula~\cite{Sterman:1994ce}; the latter projects the time evolution of gauge fields to asymptotic particle states and is cumbersome to achieve numerically. 
The resulting simulations are considerably faster relative to the dense-dense case and the continuum limit (where the independence of observables on the lattice spacing and the size of the lattice is established) is simpler to achieve as well. 

In the dilute-dense framework, it is sufficient to know $f_1$ to obtain the leading contributions to $v_{2,4,\cdots}$, the even azimuthal anisotropy coefficients. This is not the case for the odd anisotropy coefficients $v_{1,3,\cdots}$ because the leading order $f_1$ contributions preserve a momentum reflection symmetry which is manifestly violated for the odd anisotropy coefficients.  The first nonzero contribution to these odd coefficients  comes from the function $f_2$ in Eq.~\eqref{Eq:SIP}.  Because this  function appears at order $( g^2 \rho_{\rm pr}/k_\perp^2)^3$ in Eq.~\eqref{Eq:SIP}, corresponding to interactions with two projectile valence-like  color sources in either the production amplitude or its complex conjugate, it can be termed the {\it first projectile saturation correction} in the dilute-dense CGC framework. This function, at least in case of a Gaussian ensemble for the projectile,
does not contribute to the single inclusive gluon production, as it is odd in the projectile sources. 
The actual saturation correction to single inclusive production thus originates from $f_3$. 
 Efforts to compute $f_3$ analytically were discussed previously in Ref.~\cite{Balitsky:2004rr}, and more recently in Ref.~\cite{Chirilli:2015tea}; a full analytical computation is not available to date. 
 
 For the case of double inclusive production\footnote{See Eq.~\eqref{eq:Glitter} for classical n-gluon inclusive production.}, the first projectile saturation correction has two contributions; one of these is simply $f_2^2$ but there is an interference term $f_1 f_3$.  Both contributions are proportional to the (even) sixth power of the projectile density. Since currently  no complete analytical result exists for $f_3$, the complete analytical expression for the first saturation correction to double inclusive production is also unknown.    

However a breakthrough was achieved recently from the realization that the function $f_3$ is not necessary to compute the leading contributions to the odd anisotropy coefficients for two-gluon inclusive production; instead, to obtain the leading contribution, it is 
sufficient to only compute the odd piece of  $f_2$ which violates reflection symmetry~\cite{McLerran:2016snu,Kovchegov:2018jun}. With this understanding, the configuration-by-configuration parity-even and parity-odd contributions to gluon production can be expressed as~\cite{Mace:2018vwq} 
\begin{widetext}
\begin{align}
\label{eqn:dNevenodd}
	\frac{dN_g^{\rm even,\, odd}(\v{k})}{d^2k_\perp dy}\Big[\rho_{\rm pr}, \rho_{\rm t}\Big]
	=
	\frac12 \left( \frac{dN_g (\v{k})}{d^2k_\perp dy}\,\Big[\rho_{\rm pr}, \rho_{\rm t}\Big] \pm
	\frac{dN_g (-\v{k})}{d^2k_\perp dy}\Big[\rho_{\rm pr}, \rho_{\rm t}\Big] \right)\,,
\end{align}
where 
\begin{align}
	\label{even}
	\frac{d N_g^{\rm even} (\v{k})}{d^2k_\perp dy} \Big[\rho_{\rm pr}, \rho_{\rm t}\Big] &=  \frac{2}{(2\pi)^3}
		\frac{ \delta_{ij} \delta_{lm}  +  \epsilon_{ij} \epsilon_{lm} }{k^2}
		\,{\tilde \Omega}^a_{ij} (\v {k})
		\left[ {\tilde \Omega}^a_{lm} (\v {k}) \right]^\star\,,\\
	\label{odd}
	  \frac{d N_g^{\rm odd} (\v{k})}{d^2 k_\perp dy} \Big[\rho_{\rm pr}, \rho_{\rm t}  \Big]
    &=
	{ \frac{2}{(2\pi)^3} }
	{\rm Im}
	\Bigg\{
		\frac{g}{{\v{ k}}^2}
		\int \frac{d^2 l}{(2\pi)^2}
				\frac{  {\rm Sign}({\v{k}\times \v{l}}) }{l^2 |\v{k}-\v{l}|^2 }
		f^{abc}
			{\tilde \Omega}^a_{ij} (\v{l})
			{\tilde \Omega}^b_{mn} (\v{k}-\v{l})
			\left[{\tilde \Omega}^{c}_{rp} (\v{k})\right]^\star
		 \\ & \times \quad
		\left[
			\left(
			{\v{ k}}^2 \epsilon^{ij} \epsilon^{mn}
		-\v{l} \cdot (\v{k} - \v{l} )
		(\epsilon^{ij} \epsilon^{mn}+\delta^{ij} \delta^{mn})
		\right) \epsilon^{rp}+
		2 \v{k} \cdot (\v{k}-\v{l}) { \epsilon^{ij} \delta^{mn}} \delta^{rp}
		\right]
	\Bigg\} \, .\notag
\end{align}
\end{widetext}
Here $\epsilon_{ij} (\delta_{ij})$ denotes the Levi-Civita symbol (Kronecker delta), and ${\tilde \Omega}_{ij}^a(\v{k})$ is the two-dimensional Fourier transform of its coordinate space counterpart
\begin{equation}
	\Omega_{ij}^a(\v{x}) =
	- g  \left(\partial_i A^{+b}_{\rm pr}(\v{x}) \right)  \partial_j U_{ab} (\v{x})\,,
	\label{Eq:Omega}
\end{equation}
where  $A_{\rm pr}^{+ a} = \frac{1}{\partial_\perp^2 }\rho_{\rm pr} ^a(x^-,\v{x})$ with $\rho^a(x^-,\v{x})\approx \rho^a(\v{x})\delta(x^-)$.
As shown in Eq.~\eqref{Eq:U}, the adjoint Wilson line $U = U_{\rm tr}$ is the functional of the corresponding nonzero component of Lorenz gauge in the target nucleus. 
Similarly to the projectile case, here $A_{\rm t}^{-a} =  \frac{1}{\partial_\perp^2} \rho_{\rm t}^a(x^+,\v{x})$. 
As we alluded to earlier, the analytical results for the leading-order parity-even part in Eq.~\eqref{even} have been long known~\cite{Kovchegov:1998bi,Dumitru:2001ux,Blaizot:2004wu,Kovner:2012jm,Kovchegov:2012nd}, while the leading-order parity-odd contribution in Eq.~\eqref{odd} was calculated only recently~\cite{McLerran:2016snu,Kovchegov:2018jun}. 

In the classical approximation, the n-gluon inclusive multiplicity can be obtained from the above 
configuration-by-configuration result by constructing the following n-product and averaging with respect to 
the projectile and target sources:
\begin{eqnarray}
\label{eq:Glitter}
\frac{d^{n}N_g}{d^{2}k_{\perp,1}dy_{1}d^{2}k_{\perp,2}dy_{2}\cdots d^{n}k_{\perp,n}dy_{n}} &=& 
\left\langle \left\langle  
\frac{dN_g}{d^{2}k_{\perp,1}dy_{1}}  \Big[\rho_{\rm pr}, \rho_{\rm t}\Big]
\cdots \frac{dN_g}{d^{n}k_{\perp,n}dy_{n}}  \Big[\rho_{\rm pr}, \rho_{\rm t}\Big] 
 \right\rangle_{\rm pr} 
 \right\rangle_{\rm t} \, \nonumber \\
&\equiv& \int  {\cal D}\rho_{\rm t}\, {\cal D}\rho_{\rm pr}\,W[\rho_{\rm pr}]\,W[\rho_{\rm t}]\,\frac{dN_g}{d^{2}k_{\perp,1}dy_{1}}  \Big[\rho_{\rm pr}, \rho_{\rm t}\Big]
\cdots \frac{dN_g}{d^{n}k_{\perp,n}dy_{n}}  \Big[\rho_{\rm pr}, \rho_{\rm t}\Big] \, .
\end{eqnarray}
This expression for the n-gluon inclusive multiplicity in Eq.~\eqref{eq:Glitter} was proven to leading logarithms in $x$ accuracy, and for parametric rapidity separations of 
order $\Delta Y\leq 1/\alpha_s$ \cite{Dumitru:2008wn,Gelis:2008rw,Gelis:2008ad}. Even though this expression appears
 factorized before averaging over the projective and 
target sources, very nontrivial multiparticle correlations are generated after the averaging. Exact analytic expressions for these are not feasible in general due to the highly nonlinear expressions appearing in Eq.~\eqref{eqn:dNevenodd}. However, as we will demonstrate, the n-gluon inclusive multiplicity can be computed straightforwardly employing numerical methods.

In the McLerran-Venugopalan (MV) model~\cite{McLerran:1993ka,McLerran:1993ni}, the weight functionals, $W[\rho_{\rm pr, t}]$, can be expressed as 
\begin{equation}
W[\rho_{\rm pr, t}] = \exp\left(-\int \frac{d^2 q}{(2\pi)^2}\,\rho_{{\rm pr},t}^a(-q)\frac{1}{2\mu_{{\rm pr},t}^2}\,\rho_{{\rm pr},t}^a (q)\right)\,.
\label{eq:MV-weight}
\end{equation}
As noted elsewhere~\cite{Dumitru:2011vk}, this form for the weight functional is more general than the MV model. It is a good approximation to the 
results obtained from JIMWLK~\cite{JalilianMarian:1997gr,JalilianMarian:1997dw,Iancu:2000hn,Ferreiro:2001qy} renormalization group evolution of the color sources in the 
projectile and target to small $x$, where the evolution of the variance $\mu_{{\rm pr},t}^2$ with decreasing $x$ is given by the Balitsky-Kovchegov equation~\cite{Balitsky:1995ub,Kovchegov:1999yj}.

We turn now to the computation of the $v_n$ coefficients in this approach. The azimuthal angular distribution of charged particles measured in a given collision event can be expressed in terms of the $n-$th harmonic anisotropy coefficients $v_n$ and global reference angles $\Psi_{n}$ as
\begin{equation}
\frac{2\pi}{N_{\rm ch}} \frac{dN}{d\phi} = 1+ 2\sum\limits_{n=1}^{\infty} v_n \cos(n(\phi-\Psi_{n}))\,.
\end{equation}
The measurements of $v_n$ in the experiment are often performed using the event plane approach with a pair of detectors ``A" and ``B" separated by a rapidity gap. In this case, the quantity measured is defined to be 
\begin{equation}
v_n\{\text{EP}\}^{\rm expt} = \frac{\left \langle \cos(n(\phi_{A} - \Psi_{n,B})) \right \rangle}{\left\langle {\rm Res}(\Psi_{n,B})\right\rangle}\,,\\
\end{equation}
where $\phi_A$ is the azimuthal angle of the particles detected by detector ``A" and $\Psi_{n,B}$ is the $n-$th order event plane angle measured using detector ``B". The symbol $\left \langle \cdots\right \rangle$ here denotes the event average. The quantity $\text{Res}(\Psi_{n,B})=\sqrt{\left<\cos(n(\Psi_{n,B}-\Psi_{n}))^2\right>}$ is the event-plane resolution which is a measure of the dispersion between $\Psi_{n,B}$ from the reference angle $\Psi_{n}$ or the true event-plane. The value of $\text{Res}(\Psi_{n,B})$ increases with the increase of the underlying anisotropy $v_n$ but suffers a random walk suppression due to finite number of particles used in the determination of $\Psi_{n,B}$~\cite{Poskanzer:1998yz}.

When $v_{n,B}$ or ${\rm Res}(\Psi_{n,B})$ is small, an approximate linear relation holds, namely $\cos(n(\Psi_{n,B}-\Psi_{n}))\propto v_{n} \sqrt{N}$~\cite{Poskanzer:1998yz}. Therefore, one expects 
\begin{equation}
v_n\{ {\rm EP} \}^{\rm expt} = \frac{\left\langle \cos(n(\phi_{A} - \Psi_{n}))\cos(n(\Psi_{n}-\Psi_{n,B})) \right\rangle}{\sqrt{\left\langle \cos^2(n(\Psi_{n}-\Psi_{n,B}))\right\rangle }}\approx\sqrt{ \left\langle v^2_{n} \right\rangle },\\
\end{equation}
in the limit of low event plane resolution. This yields the root-mean-square value of the anisotropy coefficient;  this is also what is obtained from the experimental measurements of $v_n$ using the scalar-product (SP) or two-particle correlation methods~\cite{Luzum:2012da}:
\begin{equation}
v_n\{\text{SP}\}=\sqrt{\left\langle v_n^2 \right\rangle }\equiv \sqrt{\left\langle \cos(n\Delta\phi)\right\rangle} = \sqrt{ \left\langle v^2_n\{2\}\right\rangle }. 
\end{equation}
Since the event-plane resolutions quoted for the experimental measurements of $v_n\{\text{EP}\}$ in $p$/$d$/$^3$He+A systems are small ($<10\%$)~\cite{PHENIX:2018lia,Aidala:2018mcw}, we will approximate such measurements as $\sqrt{\left\langle v_n^2\right \rangle}$ and estimate the same in our framework.

The principal objects of our investigation are the $v_n$ anisotropy coefficients of gluons in collisions of light nuclei with heavy nuclei.  Defining 
\begin{equation}
\label{eq:vn}
	V_n (p_1, p_2)  = \frac{ \int_{p_1}^{p_2}   k_\perp dk_\perp\frac{d\phi}{2\pi} e^{i n \phi} \frac{d N (\v{k})}{d^2k dy} \Big[\rho_p, \rho_t\Big] 
  }     {   \int_{p_1}^{p_2}   k_\perp dk_\perp \frac{d\phi}{2\pi}\frac{d N (\v{k})}{d^2k dy} \Big[\rho_p, \rho_t\Big] } \,,
\end{equation}
the two-particle  anisotropy coefficients, averaged over all color charge configurations in an event and over all events, can be expressed in terms of the l.h.s. of Eq.~\eqref{eqn:dNevenodd} as 
\begin{align}
	v_n^2\{2\} (p_\perp) 
	&=   \int {\cal D} \rho_p {\cal D} \rho_t\  W[\rho_p] \ W[\rho_t]  V_n (p_\perp-\Delta/2,p_\perp+\Delta/2) V^\star_n (0, \Lambda_{UV}) \,.
\label{eq:vn-formula}
\end{align}
We consider $\Delta=0.5$ GeV bins in $p_\perp$.
A similar computation was performed previously in the dense-dense CGC framework to compute $v_2$ and $v_3$~\cite{Schenke:2015aqa}, of which the dilute-dense framework is an approximation. As noted, the latter has the significant computational advantage that analytical expressions can be written down, and more importantly, it does not require numerical evaluation of the temporal evolution of the Yang-Mills equations. The dilute-dense formalism can be employed systematically to compute higher order cumulants, as was demonstrated in the simpler toy model computation of \cite{Dusling:2017dqg,Dusling:2017aot}. We will however focus here on the two-particle $v_n$ cumulants, specifically $v_{2,3}$.

We should also comment that the aforementioned function $f_1$ in Eq.~\eqref{Eq:SIP} is fully defined by the parity-even contribution of Eq.~\eqref{even} and the momentum-odd part of the function $f_2$ 
is fully defined by the parity-odd contribution in Eq.~\eqref{odd}. One then observes straightforwardly that
the power counting in the projectile color charge density $\rho_{\rm pr}$ is different for the parity-even and parity-odd contributions. As discussed in Ref.~\cite{Mace:2018yvl}, this  observation leads to a simple scaling prediction for the dependence of the two-particle harmonics $v_2\{2\}$ and $v_3\{2\}$ on the multiplicity in proton-nucleus collisions: 
\begin{eqnarray}
  v_{2n}\{2\} \propto N_{\rm ch}^0,  \quad \quad
    v_{2n+1}\{2\} \propto N^{1/2}_{\rm ch}\,.
\end{eqnarray}

Due to the complicated structure of the integrals in Eqs.~\eqref{even} and \eqref{odd}, it is difficult to proceed further relying only on analytical computations.
However under certain further approximations, one can simplify the problem. In particular, in Ref.~\cite{ Altinoluk:2018ogz}, the authors extracted leading order results 
for the two and three particle correlations in the limit of large number of colors $N_c$ and the projectile transverse area $S_\perp$. The expressions obtained are quite transparent and clearly demonstrate the HBT and Bose-Einstein correlations originating from the quantum statistics of gluons. 

The $v_n$ anisotropies we will compute here and compare to data from light-heavy ion collisions at RHIC are those generated for gluons produced in the light-heavy ion collisions. 
If the gluon multiplicity is large enough, and the lifetime of the system is sufficiently long, strong rescattering can occur, justifying a hydrodynamic description of its spacetime evolution. Parametric estimates in the ``bottom-up" thermalization scenario~\cite{Baier:2000sb} that describe the late time scattering dynamics of  CGC generated Glasma fields in collisions of heavy nuclei suggest as much.  This is confirmed in more detailed kinetic theory estimates~\cite{Kurkela:2015qoa,Kurkela:2018wud} and are consistent with the phenomenology of the IP-Glasma+MUSIC model~\cite{Gale:2012rq}. For the smaller systems, these very estimates suggest that hydrodynamic flow is unlikely except in very rare events. However final state rescattering contributions to the azimuthal anisotropies cannot be discounted and their full description, as one varies the charged particle multiplicity, should show the transition from initial state dominance to one dominated by final state rescattering; work in this direction can be found in \cite{Greif:2017bnr,Ruggieri:2018rzi,Kurkela:2018qeb}. The AMPT event generator is an example of an event generator implementing kinetic theory with string-like initial conditions~\cite{Lin:2004en}. 

The underlying impetus for this work to ask {\it whether initial state correlations alone are sufficient} to explain the magnitude and systematics of the data on $v_n$ anisotropy coefficients measured in light-heavy ion collisions at RHIC and the LHC. To go further and include rescattering corrections would introduce another layer of systematic uncertainty to uncertainties in our understanding of the initial state. This is because, despite much progress, there still are significant uncertainties in the kinetic theory approaches. 

For the collisions of large nuclei, our approach clearly produces anisotropies that are too small~\cite{Schenke:2015aqa} necessitating including final state rescattering contributions. This can be understood as lending theoretical support and an initial condition framework for the hydrodynamical picture in central collisions of large nuclei.  We will see below that this is not the case for the light-heavy ion collisions. While final state re-scattering may occur, and should be quantified, its effects do not appear to be essential, within theory uncertainties, to describe data -- at least published data in the centrality windows measured by the RHIC experiments. 

A major uncertainty bedeviling all theoretical approaches is the impact of fragmentation of soft partons into hadrons. For high $p_\perp$, one may use fragmentation functions but even in that case there is considerable uncertainty in how multiparton correlations hadronize. At low $p_\perp$, string-like models of fragmentation are tuned to data and successfully describe many of its features
~\cite{Sjostrand:2014zea,Pierog:2013ria,Bellm:2017bvx}. In hydrodynamical approaches, one employs a prescription whose first principles justification is thin; examples are the Cooper-Frye~\cite{Cooper:1974mv} and Sinyukov prescriptions~\cite{Sinyukov:2002if} -- for short lifetimes, uncertainties in these are magnified. Alternately, one may match the hydro output to a hadron transport afterburner as is the case in Ref.~\cite{Shen:2014vra}; to the best of our knowledge, these uncertainties have not been quantified for the smaller systems, in particular for high multiplicity events. 

In Refs.~\cite{Schenke:2016lrs,Schenke:2018hbz} gluons from the dense-dense collision (IP-Glasma) were converted to hadrons using the PYTHIA event generator~\cite{Sjostrand:2014zea}: while reasonable agreement is achieved, and a mass ordering of $v_2$ believed previously to be a signature of hydrodynamic flow was seen, one is nevertheless implementing a fragmentation prescription whose applicability to high multiplicity events is not guaranteed. In the CGC+PYTHIA~\cite{Schenke:2016lrs,Schenke:2018hbz} studies, it was observed that the $p_\perp$ spectra of 
charged pions is very similar to that of gluons at low $p_\perp$; we will show here that the shape of the spectrum for low $p_\perp$ for gluons in the dilute-dense framework well approximates that for pions in deuteron-gold collisions. This does not mitigate the fundamental systematic uncertainty of hadronization afflicting all theory approaches but does indicate that our approach of comparing gluon $v_2$ spectra to data at low $p_\perp$ is no worse than any of the other prescriptions used for theory-data comparisons in the literature. One further hopes that some of the fragmentation uncertainties will cancel in the ratio taken in Eq.~\eqref{eq:vn-formula}.

\section{Numerical framework for the event-by-event dilute-dense CGC computations}

In this section, we will provide details of the numerical computation of gluon production and correlations in Refs.~\cite{Mace:2018vwq,Mace:2018yvl} and in the additional computations carried out in this paper. Our objective specifically is to spell out all the ingredients that go into the numerical computation of the even and odd parity contributions to Eq.~\eqref{eqn:dNevenodd} and the subsequent comparision of Eq.~\eqref{eq:vn-formula} to data. This requires that one first compute the spatial distribution of color charge densities in the projectile and target nuclei; for the latter, one requires further the computation of the corresponding path ordered light-like Wilson line. To ensure gauge covariance of this quantity, it is essential to employ the discretization techniques of lattice gauge theory~\cite{Krasnitz:1998ns,Krasnitz:1999wc} that were also employed in the IP-Glasma model~\cite{Schenke:2012wb,Schenke:2012fw}. We next compute the function $\Omega(\mathbf{x})$ 
in Eq.~\eqref{Eq:Omega} which then allows for the direct computation of Eqs.~\eqref{even} and~\eqref{odd}.
          
\subsection{Initialization of color charge densities}
We will describe here how one constructs the event-by-event color charge densities that are necessary to compute Eqs.~\eqref{even} and~\eqref{odd}. This is achieved as follows:

{\it i) Nucleon position sampling}:  
For a single event, we sample the nucleon positions for the projectile and target nucleus using known wavefunctions for the nuclei. The procedure we follow is identical to the Monte-Carlo Glauber modeling described in Ref.~\cite{Miller:2007ri}. For the deuteron, we employ the Hulth\'{e}n wave function parameterization of \cite{Miller:2007ri} that allows us to model the spatial proton and neutron distribution in the deuteron. For the $^3$He projectile, we use configurations from Refs.~\cite{Nagle:2013lja,Loizides:2014vua}, which were generated using ab-initio Green's function Monte-Carlo methods. The nucleons in the gold nuclear target are sampled from a three-parameter Fermi distribution, using the parameterization of Ref.~\cite{Loizides:2014vua}. Configurations where there are no participant nucleons, whereby all nucleons in the colliding nuclei are separated in the transverse plane by distances greater than $\sqrt{\sigma_{NN}/\pi}$, are rejected. We take the nucleon-nucleon cross-section value\footnote{In Ref.~\cite{Mace:2018yvl}, for the LHC $\sqrt{s}=5.02~\text{TeV}$/nucleon, we used $\sigma_{NN}=67.6~\text{mb}$~\cite{Loizides:2017ack}.} $\sqrt{s}=200~\text{GeV}$/nucleon~\cite{Loizides:2017ack} $\sigma_{NN}=42~\text{mb}$ for RHIC. 
 At this point, an impact parameter is also specified from a uniform distribution.

{\it ii) Impact parameter dependent saturation momentum}:
    We next determine the color charge density for each nucleus using a procedure analogous to that employed in the IP-Glasma model~\cite{Schenke:2012wb,Schenke:2012fw}. To achieve this, we  first employ the IP-Sat model~\cite{Bartels:2002cj,Kowalski:2003hm} to determine the saturation scale $Q_s(x,\v{x})$ for each nucleon; here $\v{x}$ is the impact parameter and $x$ is the Bjorken-$x$ variable\footnote{Anisotropic subnucleon scale spatial color charge fluctuations~\cite{Mantysaari:2016ykx} are not included since a previous study showed that these have little or no impact on initial state momentum correlations~\cite{Schenke:2015aqa,Kovner:2018fxj}. In contrast, when initial state models including such fluctuations are matched on to hydrodynamics, the enhanced spatial ellipticity from these fluctuations leads to significant increases in the anisotropy coefficients~\cite{Mantysaari:2017cni}.} 

 The IP-Sat model parameterizes the nuclear DIS dipole scattering amplitude in terms of the strong coupling constant\footnote{For simplicity, we do not include the running of the coupling, the effect of which was examined previously in the IP-Glasma model~\cite{Schenke:2013dpa}.For ratios of the multiplicity in A+A collisions, different running coupling prescriptions give similar results for low 
 $N_{\rm part}$, where $N_{\rm part}$ is the number of participants. For large $N_{\rm part}$ the differences between prescriptions are $\sim 20$\%.} $\as$, the gluon distribution $xg(x,\mu^2)$, and the nucleon impact parameter profile (or thickness function) $T_p$ as~\cite{Kowalski:2007rw}
    \begin{eqnarray}
    \mathcal{N}_A(x,\v{r},\v{x})=1-e^{-\frac{\pi^2}{2N_c}|\v{r}|^2 \as xg(x,\mu^2)\sum_{i=1}^A T_p(\v{x}-\mathbf{x}_\perp^i)}\,.
    \end{eqnarray}
 As in Ref.~\cite{Schenke:2013dpa}, $T_p$ is a Gaussian distribution with width $B_G$, which provides a measure of the root mean square gluon radius of a nucleon. We remove the unphysical tails of these profiles for $T_p<T_p(|\v{x}|_{\rm max}) \equiv T_p^{\rm min}$; to be conservative, we take $|\v{x}|_{\rm max}=3~\text{fm}$, which coincides with approximately five times the average gluon radius of the proton~\cite{Caldwell:2010zza}. We have checked that our results do not change appreciably by taking a more restrictive value of $|\v{x}|=1.2~\text{fm}$.

  The gluon distribution in the proton is evolved using small x double log DGLAP evolution with the scale $\mu^2=4/|\v{r}|^2+\mu_0^2$~\cite{Rezaeian:2012ji}. The parameter values  $B_G=4~\text{GeV}^{-2}$ and $\mu_0^2=1.51~\text{GeV}^2$ were previously determined from IP-Sat fits to HERA DIS data~\cite{Rezaeian:2012ji}. The nuclear saturation scale $Q_s^A$ is then determined by solving $\mathcal{N}\left(|\v{r}|=\frac{1}{\Qs^A} \right)=1-e^{-1/2}$. Since $\Qs^A$ depends on both $x$ and $\v{x}$, we must simultaneously solve the relation $x= Q_s^A(x,\v{x})/\sqrt{s}$ to determine $Q_s^A(x,\v{x})$ in a nucleus at a given center of mass energy. Following these steps, one can determine $Q_s^A(x,\v{x})$ for each nuclear configuration in both the projectile and the target independently.   
  
{\it iii) Saturation momentum fluctuations}:
In addition to the fluctuations induced by nucleon position fluctuations, the nuclear saturation scale $\Qs^A$ can fluctuate due to event-by-event fluctuations in the number and correlations of gluons in the nucleus as one evolves the nucleus to small x. These are not included naturally in the IP-Sat model; while this model has the virtue that small x dynamics is constrained by HERA DIS data on the proton, it does not include all the 
 complexity of small x evolution in the first principles JIMWLK framework. For $\Qs^A$, there are event-by-event Gaussian fluctuations of $\ln(\Qs^A)$~\cite{Iancu:2001md,Iancu:2004es,Marquet:2006xm,Iancu:2007st,Kovner:2018azs}. In the IP-Sat approach, the variance of these fluctuations, which we denote as $\sigma$, is fixed by fits to multiplicity distributions. It was observed in \cite{Schenke:2013dpa} that color charge fluctuations {\it a la} IP-Sat alone are insufficient to reproduce the very high multiplicity tails of these distributions and that fluctuations in $\Qs^A$ are 
 necessary to fit these~\cite{McLerran:2015qxa}. Conversely, a fit the multiplicity distribution constrains the variance $\sigma$ of $\Qs$ fluctuations. We will return to this issue when we discuss 
 multiplicity distributions later in the manuscript.
      
Once a final local value of $\Qs^A(x,\v{x})$ for a nuclear configuration is determined by including fluctuations in the nucleon positions in the IP-Sat model, and is further Gaussian distributed around this value with the variance $\sigma$,  we relate each such value of $\Qs^A$ to the color charge squared per unit area $g^2\mu_{\rm pr, t}(x,\v{x})$ in both the projectile and target nucleus via a 
constant nonperturbative coefficient
 \begin{eqnarray}
\mathcal{K} \equiv \frac{Q_s^A(x,\v{x})}{g^2 \mu_{\rm pr, t}(x,\v{x})}\,. 
\label{eqn:prop_kappa}
 \end{eqnarray}
This constant is a free parameter in our model which has only been estimated before for the MV model\footnote{Since the weight functional $W[\rho]$ is a nonperturbative density matrix, the proportionality factor  $\mathcal{K}$ should be thought of as the effective gauge coupling between the number density $(Q_s^A)^2$ and the color charge squared per unit area $\mu^2$. The additional factors of $g$ in Eqs.~\eqref{eqn:prop_kappa} and \eqref{eqn:rhocorrcont} are effectively subsumed in $\mathcal{K}$. Indeed, since $g$ can be scaled out of the lattice computation entirely, it can equivalently be set to unity. In other words, $\mathcal{K}$ relates the physical scale $(Q_s^A)^2$ extracted from DIS experiments to quantities computed on the lattice expressed as functions of $g^2\mu^2$. A careful discussion of the relation between the two is given in Ref.~\cite{Lappi:2007ku} for the specific case of the MV model. It is somewhat sensitive to the value of $N_y$ and the infrared scale $m$ defined there but converges at $\mathcal{K}_{\rm MV}=0.8$. The latter however does not include nucleon position or $Q_s$ fluctuations.}; we will use here $\mathcal{K}=0.5$~\cite{Mace:2018vwq,Mace:2018yvl}. 

{\it iv) Color charge density fluctuations}:
      The color charge squared per unit area $g^2\mu^2_{\rm pr, t}(x,\v{x})$, thereby extracted, acts as the width of the color charge density fluctuations. Namely, we sample the color charge densities $\rho_{\rm pr, t}^a(x^\mp,\v{x})$,using the Gaussian distribution~\cite{McLerran:1993ni,McLerran:1993ka},
\begin{eqnarray}
\langle \rho_{\rm pr, t}^a(x^\mp,\v{x}) \rho_{\rm pr, t}^b(y^\mp,\v{y}) \rangle = g^2\mu^2_{\rm pr, t}(x,\v{x}) \delta^{ab} \delta(x^\mp-y^\mp) \delta^2(\v{x}-\v{y})\,,
\label{eqn:rhocorrcont}
\end{eqnarray}
where $a=1,...N_c^2-1$. These color charge densities in the projectile and target are essential to determine all required quantities in Eqs.~\eqref{even} and \eqref{odd}. 

{\it v) Lorenz gauge field of projectile and target}:
Now that we have specified the spatial distribution of color charge densities $\rho_{\rm pr, t}^a(x^\mp,\v{x})$, we compute the gauge field $A^+$ for the projectile moving in the $x^+$ direction and the corresponding gauge field $A^-$ for the target moving in the $x^-$ direction in their respective Lorenz gauges before the collision: 
\begin{eqnarray}
A^{\pm}_{\rm pr, t} (x^{\mp},\v{x}) = \frac{  \rho_{\rm pr, t}(x^{\mp},\v{x})}{{\partial}_\perp^2+m^2}\,.
\label{eqn:poisson_cont}
\end{eqnarray}
We have introduced here a nonperturbative regulator mass $m$ to remove zero mode divergences that arise from the solution of the Poisson equation. We fix $m$ by minimizing deviations\footnote{For simplicity, we take this parameter to be the same for both nuclei.} from the observed multiplicity distribution; in Refs.~\cite{Mace:2018vwq,Mace:2018yvl} this quantity was determined to be 0.3 GeV. We have checked that our results do not vary appreciably when $m$ is varied by $\pm 0.1$ GeV.
The Lorenz gauge fields $A^\pm$ thus computed are used to determine Eqs.~\eqref{Eq:U} and \eqref{Eq:Omega}. 

It is important to note that even though the color source densities are Gaussian distributed, this does not imply that the corresponding gauge field $A^\pm$ are Gaussian distributed as well. On very general grounds, a solution of the  Poisson equation in Eq.~\eqref{eqn:poisson_cont}
smears the profile of $\rho_{\rm pr, t}$, such that the tails of the functions  $A^\pm$ describing the soft gluon fields fall off exponentially--slower than the Gaussian profiles of the valence charges.

\begin{table}[t]
\begin{tabular}{|l|l|l|l|l|l|l|}
\hline
Parameter & $m$ & $\mathcal{K}$ & $\sigma$  & $B_G$ &  $\mu_0^2$ & $\sigma_{NN}$ \\
\hline
Constrained by  & d-A & d-A & d-A& DIS  & DIS & p-p($\bar{\rm p}$)\\
\hline
Value & 0.3 GeV & 0.5 & 0.5& 4 GeV$^{-2}$  & 1.51 GeV$^2$  & 42 mb \\
\hline
\end{tabular}
\caption{
\label{T:Par}
Table of the parameters used in the present calculation with the indication of the experiments whereby they are constrained. See text for discussion.} 
\end{table}

In Table~\ref{T:Par}, we summarize the parameters used in the present calculations and how these parameters are constrained. As noted, $\sigma_{\rm NN}$ is determined from $p$+$p$ collisions at RHIC energies; $B_G$ and $\mu_0^2$ are determined from fits to the HERA DIS data on electron-proton scattering. The free parameters therefore are $m$, ${\mathcal K}$ and $\sigma$. One parameter we have not listed separately is the minimum value $T_p^{\rm min}$ of the Gaussian nucleon profile function; as noted, we have checked that our results are insensitive to changes in this quantity. As we will soon discuss, their values are constrained by the RHIC $p$+$p$ and $d$+Au multiplicity distributions.

\subsection{Lattice implementation}
To numerically compute the  gluon gauge field distributions of the projectile and target nuclei in each event before the collision, we will employ a lattice discretization of the Yang-Mills equations~\cite{Krasnitz:1998ns,Krasnitz:1999wc,Lappi:2007ku}. We consider isotropic lattices of transverse extent $L_\perp$, which have $N_\perp$ sites with lattice spacing  $a_\perp$ (in units of fermi) for both the projectile and target fields such that $L_\perp = N_\perp a_\perp$. As discussed in \myref\cite{Lappi:2007ku}, it is necessary to take into account the path ordering in Eq.~\eqref{Eq:U}. Since the path ordering is only on the target side in the dilute-dense approximation, we will keep the $x^-$ dependence on the target nucleus color charge density, and neglect the $x^+$ dependence in the projectile color density.

The effective width of the target is characterized by the variable $x^+$. The 
 $x^+$-delta function in Eq.~\eqref{eqn:rhocorrcont} demonstrates that the correlations of the target sources are local not only in the 
 transverse plane but also in their longitudinal extent. Following Ref.~\cite{Lappi:2007ku}, we discretize the  $x^+$-delta function in Eq.~\eqref{eqn:rhocorrcont} as 
\begin{eqnarray}
\langle \rho_{\rm t}^a(i_y, \v{i}) \rho_{\rm t}^b(j_y, \v{j}) \rangle_{\rm t} =\delta_{i_yj_y} \,\delta^{ab}\, \frac {   \delta_{\v{i} \v{j}}  }{a_\perp^2}\, \frac{ g^2\mu^2_{\rm t}(\v{i})}{N_y}\,,
\label{eqn:rhocorrlat}
\end{eqnarray}
where $N_y$ are the number of the target slices in the longitudinal $x^+$ direction and $i_y,j_y=1,...,N_y$ label these slices. 
The transverse coordinates are related to their discretized partners according to $\v{x}=a_\perp  \v{i}$ with $i_1, i_2 =1, ..., N_\perp$. 
We use periodic boundary conditions in the transverse directions.  
We find that our results converge to the continuum limit of $N_y\to \infty$ well before our chosen value of $N_y=100$~\cite{Lappi:2007ku}.

Since the Wilson lines associated with the projectile color sources are expanded into a power series, and the corresponding path ordering
is analytically accounted for, it is sufficient to consider one rapidity slice for the projectile.

We will solve the Poisson equation for the fields $A^{\pm}_{\rm pr, t}$ in Eq.~\eqref{eqn:poisson_cont} by Fourier transforming the expression to momentum space. We first fast Fourier transform the projectile/target color sources $\rho_{\rm pr, t}(x^{\mp}, \v{x})$ to obtain $\rho_{\rm pr, t}(x^{\mp},\v{p})$. Eq.~\eqref{eqn:poisson_cont} can be represented on the momentum space (reciprocal) lattice (denoted by momentum label $\v{I}$ with  $I_1,I_2=0,...,N_\perp-1$) in the form~\cite{Press:2007:NRE:1403886}, 
\begin{eqnarray}
A^{\pm}_{\rm pr, t}(i_y,\v{I})= 
\frac{\rho_{\rm pr, t}(i_y,\v{I}) }{\frac{2}{a_\perp^2}\left[ {\rm cos}\left(\frac{2\pi I_1}{N_\perp}\right) +{\rm cos}\left(\frac{2\pi I_2}{N_\perp}\right) -2 \right]+m^2}\,.
\label{eq:NP}
\end{eqnarray}
For the projectile nucleus, as noted, the expression is taken to be independent of $i_y$ while it depends on $i_y$ for the target nucleus. The mass-independent piece of the denominator in this expression, being an eigenvalue of the $\partial_\perp^2$ operator, is nothing other than the physical momentum squared defined on a periodic 2-dimensional lattice, 
\begin{eqnarray}
k^2_{\v{I}} = -\frac{2}{a_\perp^2}\left[ {\rm cos}\left(\frac{2\pi I_1}{N_\perp}\right) +{\rm cos}\left(\frac{2\pi I_2}{N_\perp}\right) -2 \right]\,.
\label{eq:mom2}
\end{eqnarray}

One can similarly  define the vector momentum on the lattice\footnote{We employ the central finite difference scheme for all derivatives, whereby 
$\partial f(x_1, x_2) /\partial x_1 \to (f(i+1,j)-f(i-1,j))/2 a_\perp $. 
}  as the eigenvalue operator of $\partial_i$ to be 
\begin{eqnarray}
\vec{k}_{\v{I}} = \frac{1}{a_\perp}
\left(
\sin \left( \frac{2\pi I_1}{N_\perp} \right) , 
\sin \left( \frac{2\pi I_2}{N_\perp} \right)  
\right)\,.
\label{eq:mom}
\end{eqnarray}
From Eq.~\eqref{eq:mom}, one concludes that the maximum momentum on the lattice is 
governed by the inverse spatial resolution $\frac{1}{a_\perp}$ and the momentum discretization step is given by  $2\pi/L_\perp$.

The expression in Eq.~\eqref{eq:NP} is then numerically fast Fourier  transformed back to coordinate space to obtain $A^{\pm}_{\rm pr, t}(\v{x})$.

For the target, one additionally needs to compute the target Wilson line in Eq.~\eqref{Eq:U} using the path ordering discussed previously. On the lattice it reads 
\begin{eqnarray}
U(\v{i})=\prod_{i_y=1}^{N_y}\exp \left[- ig A^-_{\rm t}(i_y,\v{i})\right]\,.
\end{eqnarray}

In the numerical computations, transverse lattice grids with areas $L_\perp^2=16^2-32^2~\text{fm}^2$, having $N_\perp^2=512^2-2048^2$ sites respectively are investigated to ensure that the continuum limit is reached in both the infinite volume ($a_\perp$ fixed, $N_\perp\to \infty$) and continuum limits ($L_\perp=N_\perp a_\perp$ fixed, $a_\perp\to0$ and $N_\perp\to\infty$).

\subsection{Gluon production}

The next step towards computing gluon production is to use the lattice discretized gauge fields and Wilson lines to construct $\Omega_{ij}(\v{x})$ in Eq.~\eqref{Eq:Omega}. Since the eigenvalues of operators $\partial_i$ and $\partial_\perp^2$ of Eqs.~\eqref{eq:mom} and \eqref{eq:mom2} were computed numerically using central finite difference derivatives, for consistency 
we will compute the spatial derivatives in Eq.~\eqref{Eq:Omega} using the same technique.
With $\Omega_{ij}(\v{x})$ thus obtained, we can again apply a numerical fast Fourier transform to evaluate Eqs.~\eqref{even} and~\eqref{odd}. Note that for the parity-odd piece in Eq.~\eqref{odd}, the momentum integral over $l$ has to be performed; it is computed using standard quadrature integration over all points in the momentum reciprocal-lattice. 

We can now calculate Eq.~\eqref{eqn:dNevenodd} for a given configuration of $\rho_{\rm pr}$ and $\rho_{\rm t}$. The event averaged single inclusive multiplicity is determined by integrating over $d^2k$ -- thus only the even part given in Eq.~\eqref{even} contributes.  Multiparticle correlations are obtained by taking products of Eq.~\eqref{eqn:dNevenodd} as suggested by 
Eq.~\eqref{eq:Glitter} for each configuration of $\rho_{\rm pr}$ and $\rho_{\rm t}$ and averaging over all configurations. The corresponding double, triple and n-inclusive distributions generate nontrivial quantum correlations~\cite{Dumitru:2008wn,Gelis:2008rw,Gelis:2008ad,Dusling:2009ar,Gelis:2009wh} for which the color averaging procedure we have described is absolutely essential.

For the specific case of double inclusive distributions in the dilute-dense framework, there have been a number of analytic studies that have examined these correlations more closely identifying in particular Bose enhancement~\cite{Kovner:2012jm,Altinoluk:2015uaa,Kovner:2018azs,Martinez:2018tuf} and Hanbury-Brown--Twiss (HBT)~\cite{Kovchegov:2012nd,Altinoluk:2015eka,Kovner:2017ssr,Martinez:2018tuf} contributions. It is important to note that these quantum interference effects are contained in our numerical simulations. We will revisit multiparticle correlations in more detail in the discussion section. 

Calculating multiparticle observables as in Eq.~\eqref{eq:Glitter} involves  sampling  over a large number of color charge configurations.  In the CGC EFT, this second step is crucial in order to ensure gauge invariance~\cite{McLerran:1993ni,McLerran:1993ka}. In performing this average, as outlined in the previous subsections, one is taking into account fluctuations in the nucleon positions, in the saturation momenta, as well as in the color charge densities of the target and projectile. We sample on the order of 10,000 configurations for a single system to compute the n-gluon multiplicity 
distribution~\footnote{This large number of the configuration is required to access the high multiplicity tail of gluon production using the our brute force numerical method. A more efficient procedure would be to use a biased averaging which is currently under development~\cite{Dumitru:2018iko}.}. As discussed previously in Ref.~\cite{Gelis:2009wh}, and recently revisited in Ref.~\cite{Kovner:2018azs}, this n-gluon distribution is a negative binomial distribution in the MV model; due to the presence of impact parameter and saturation scale fluctuations, our computed multiplicity distribution is a convolution of negative binomial distributions. 

In order to compare to experimentally measured multiplicity distributions, we will assume a parton-hadron duality between gluons and charged hadrons, as discussed in the previous section. As we will discuss shortly,  the comparison between the computed n-gluon multiplicity distribution and the experimentally measured proton-proton $p$+$p$ and $d$+Au multiplicity distributions at RHIC can be used to constrain the nonperturbative parameters $m$, $\sigma$, $\mathcal{K}$ and $T_p^{\rm min}$ in our framework. 

\section{Results}

\begin{figure*}[t!]
    \centering
        \begin{subfigure}[t]{0.45\linewidth}
        \centering
\includegraphics[width=\linewidth]{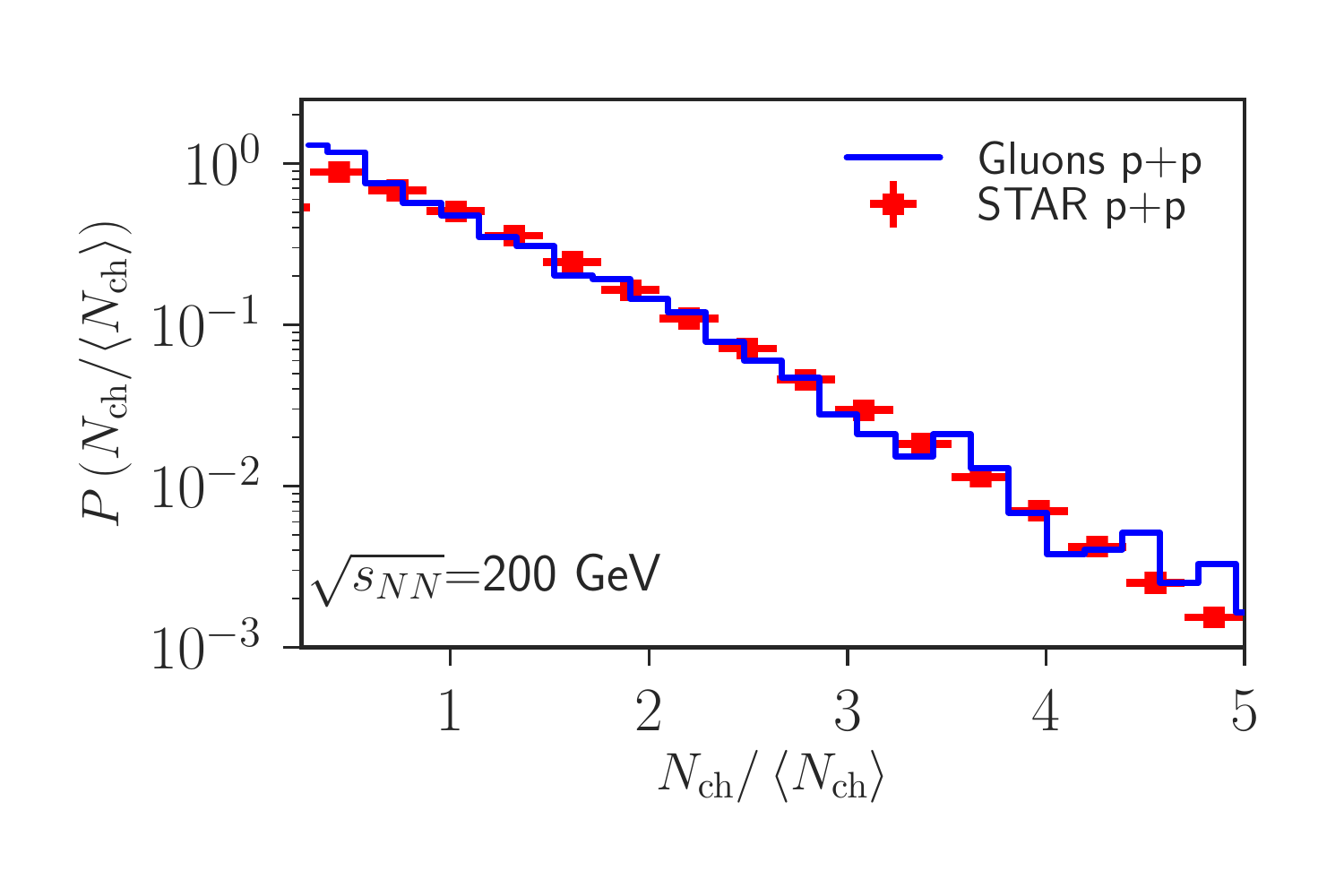}
\caption{}
\label{fig:pp_mult}
    \end{subfigure}
    \begin{subfigure}[t]{0.45\linewidth}
        \centering
\includegraphics[width=\linewidth]{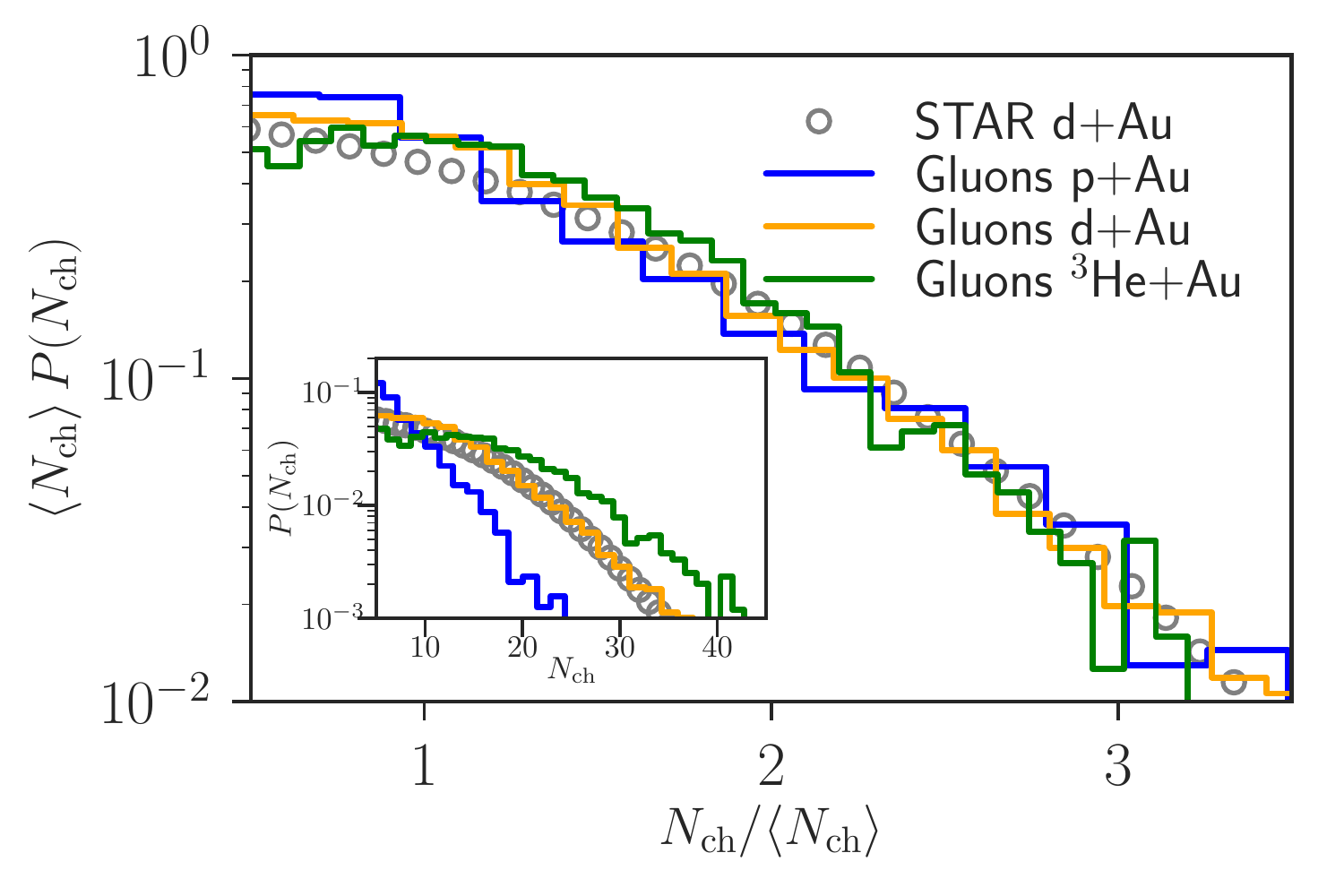}
\caption{}
\label{fig:all_mults}
    \end{subfigure}
    \caption{(a) Gluon multiplicity distribution in $p$+$p$ collisions from the dilute-dense CGC EFT as a function of $N_g/\langle N_g\rangle $ compared to the charged hadron multiplicity distribution for $N_{\rm ch}/\langle N_{\rm ch} \rangle $ from the STAR collaboration at RHIC~\cite{Abelev:2008ab}. (b) Gluon multiplicity distributions for $p$/$d$/$^3$He+Au in the dilute-dense CGC EFT. The three systems are shown to satisfy KNO scaling~\cite{Koba:1972ng}; the multiplicity distributions for the three systems is shown in the inset. Only data for the $d$+Au multiplicity distribution is available at present~\cite{Abelev:2008ab}.}
\end{figure*}

A key feature of our model is that n-gluon inclusive multiplicity distribution computed using Eq.~\eqref{eq:Glitter} can be fit to data on the charged particle multiplicity distribution in proton-proton\footnote{In applying the dilute-dense CGC framework to the former, for any given configuration, the proton with the larger $\Qs$ locally is  taken to be the target proton.} and light-heavy ion collisions at RHIC.   In Fig.~\ref{fig:pp_mult}, we see that STAR data on $P(N_{\rm ch}/\langle N_{\rm ch}\rangle)$ as a function of $N_{\rm ch}/\langle N_{\rm ch}\rangle$ is well 
reproduced, in a wide range in $N_{\rm ch}/\langle N_{\rm ch}\rangle$, by the n-gluon distribution plotted as function of $N_g/\langle N_g\rangle$. The free parameters are fixed to be the values shown in Table~\ref{T:Par}, the exception being that for this $p$+$p$ case $\sigma=0.1$. Good agreement is also found for $\sigma=0$; this indicates that $Q_s$ fluctuations are not necessary to describe the data in $p$+$p$ collisions and that color charge fluctuations in the colliding protons are sufficient to generate the n-hadron correlations seen in the data. 

The free  parameters $m$, $\sigma$, $\mathcal{K}$ and $T_p^{\rm min}$ are correlated with one another; generically, increasing $T_p^{\rm min}$ or decreasing $m$ increases the contributions 
to gluon production respectively from the tails in the impact parameter and transverse momentum distributions. Albeit this is a small effect, its impact is seen in low multiplicity events and affects the normalization of $P_n$; we therefore neglect very low multiplicity events and renormalize appropriately. To compensate for the rapid corresponding decrease of $P_n$ with $n$, one has to increase $\sigma$ to fit the experimental distribution. Thus even the ``free" parameters are strongly constrained. 

The take away message from this data-theory comparison is that color charge fluctuations within the colliding protons are quite important to reproduce the measured multiplicity distribution at RHIC. This was also seen to be the case in previous comparisons in this framework to data at the LHC although in that case it was seen that additional $Q_s$ fluctuations were needed in the highest multiplicity events with $N_{\rm ch} / \langle N_{\rm  ch}\rangle > 3$~\cite{Schenke:2013dpa}. 

In Fig.~\ref{fig:all_mults}, we show results for the multiplicity distributions for light-heavy ion collisions at RHIC. The published data are for $d$+Au collisions from the STAR collaboration~\cite{Abelev:2008ab}, and the model parameters are fixed in comparison to these. The values for the model parameters are given in Table~\ref{T:Par}; the only change relative to $p$+$p$ collisions is that $Q_s$ fluctuations, in addition to color charge fluctuations, are essential to obtain good agreement with the $d$+Au data. With these fixed, we show results for $p$+Au and 
$^3$He+Au collisions; these are shown separately in the inset of Fig.~\ref{fig:all_mults}.  Fig.~\ref{fig:all_mults} also shows strikingly that all three systems satisfy Koba-Nielsen-Olesen (KNO) scaling~\cite{Koba:1972ng}. The importance of KNO fluctuations in describing the RHIC light-heavy ion collision data was first emphasized in \cite{Dumitru:2012yr}.

In future, data from the RHIC collaborations for the multiplicity distributions in these systems can further constrain the parameters of our model. In the absence of published data for the multiplicity distributions in $p$+Au and $^3$He+Au collisions, the centrality selections for our results for these systems are chosen employing the respective gluon multiplicity distributions. We note that the PHENIX collaboration has presented results for the  average multiplicity in the 0-5\% multiplicity class around mid-rapidity~\cite{Adare:2018toe} for all three systems. We find that our values for the average gluon multiplicity to be larger than the PHENIX values but are consistent with those quoted by STAR.

\begin{figure}
\includegraphics[width=0.45\linewidth]{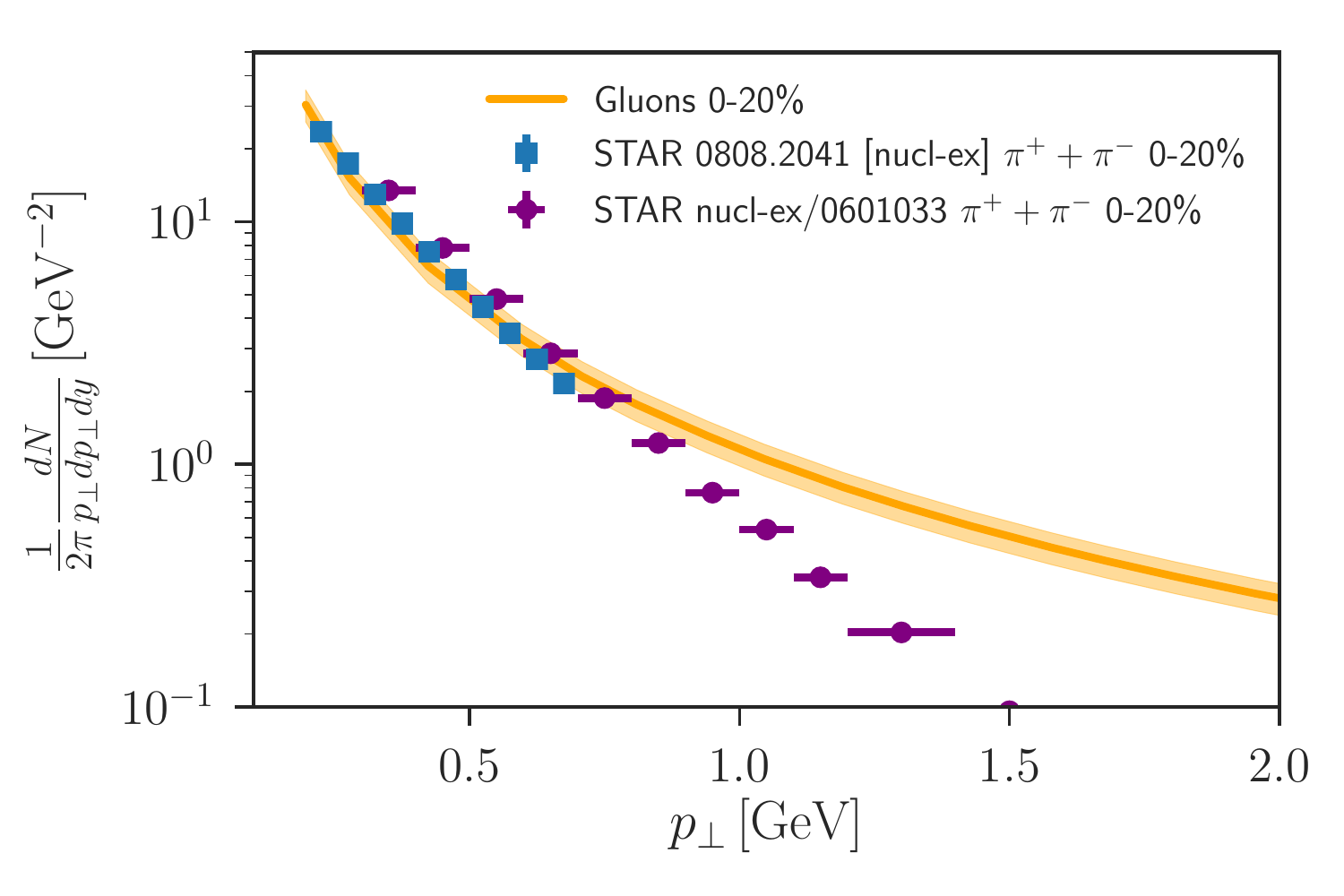}
\caption{Comparison of the gluon single inclusive distribution in comparison with pions measured by the STAR experiment in $d$+Au collisions~\cite{Adams:2006nd,Abelev:2008ab}; see text for related discussion.}
\label{fig:single_inclusive}
\end{figure}

In Fig.~\ref{fig:single_inclusive}, we compare the single inclusive $p_\perp$ distribution for gluons in our model to the sum of the charged pion single inclusive $p_\perp$ distributions in the $0$-$20$\% centrality class of $d$+Au collisions. We have normalized the gluon $p_\perp$ distribution to the lowest $p_\perp$ point in data;  the corresponding normalization factor for $g=1$ in the numerical simulations is $8.5$. With this fixed, the shape of the gluon $p_\perp$ distribution at low $p_\perp$ is in reasonable agreement with that of charged pions for low $p_\perp$. However as is clearly visible, the gluon $p_\perp$ distribution strongly deviates from the measured pion distributions for $p_\perp\geq 1$ GeV. This behavior for larger $p_\perp$ is typical for gluon distributions which are significantly corrected in the process of fragmentation. This was also seen in the study previously performed in the dense-dense framework \cite{Schenke:2013dpa} for minimum bias $p$+Pb collisions at the LHC; agreement with data was shown to be significantly improved after employing the so-called KKP fragmentation function~\cite{Kniehl:2000fe}. The sole point of our comparison is to note that the shape of the $p_\perp$ distribution in the numerator and denominator of $v_{2,3}(p_\perp)$ is similar to that of the data at low $p_\perp$, modulo the overall normalization factor that cancels between numerator and denominator.

 As noted previously, there are tremendous uncertainties in how to perform the conversion from gluons to hadrons at low $p_\perp$ that are generic to all models of low $p_\perp$ physics and ours is no exception. While some of these approaches use ``tuned" prescriptions, there is no guarantee that these are robust in rare events. One hopes that in ratios of distributions, as exemplified by multiplicity distributions of $N_g/\langle N_g \rangle $ compared to $N_{\rm ch}/\langle N_{\rm ch} \rangle $, and in $v_n$, such uncertainties cancel. A further caveat is that in general the two-particle anisotropy coefficients  are proportional to the double inclusive gluon distributions defined in Eq.~\eqref{eq:Glitter}; little is known about the fragmentation of these double inclusive gluon distributions to double inclusive hadron distributions even at high $p_\perp$.  One approximation is to replace the corresponding two-gluon fragmentation functions with the product of two single particle fragmentation functions~\cite{Dusling:2013qoz}.  We will assume henceforth that the anisotropy coefficients of gluons can be compared to those of charged hadrons; resolving the uncertainties inherent in this, and other prescriptions, is beyond the scope of this work. 

\begin{figure}
\includegraphics[width=0.95\linewidth]{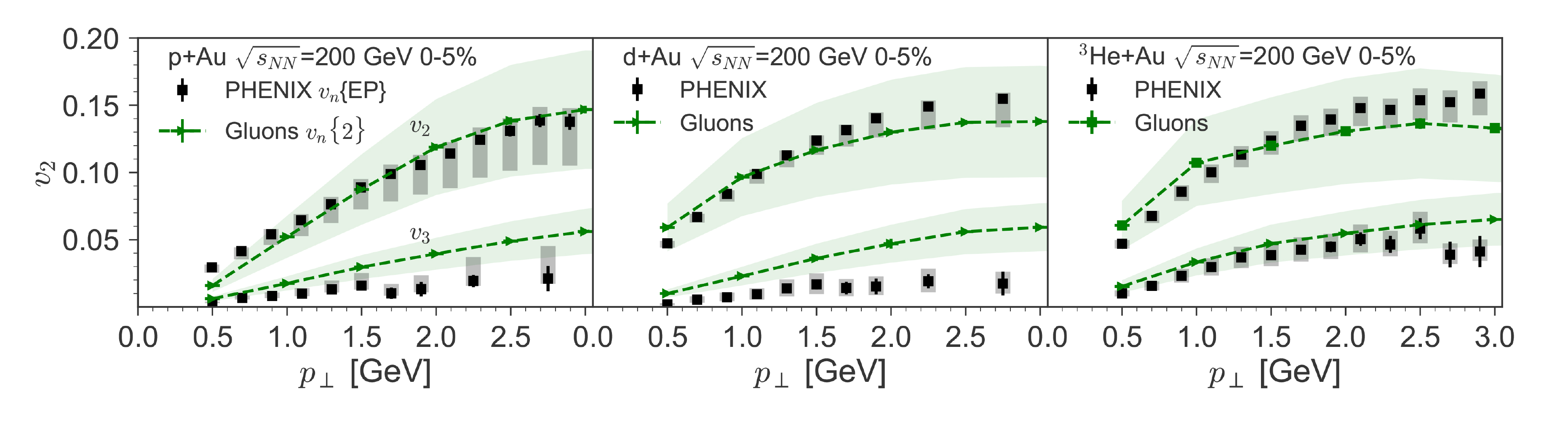}
\caption{Comparison of 0-5\% multiplicity class $v_n\{2\}(p_\perp)$ for $n=2,3$ for gluons~\cite{Mace:2018vwq}, compared to recent PHENIX results on charged hadrons~\cite{PHENIX:2018lia,Aidala:2018mcw}. The light colored (in green) bands correspond to a 30\% theory uncertainty.}
\label{fig:v2_phenix_comp}
\end{figure}

In Fig.~\ref{fig:v2_phenix_comp}, we show a comparison of $v_{2,3}\{2\}$ computed for gluons using Eq.~\eqref{eq:vn-formula} to the PHENIX data at $\sqrt{s}=200$ GeV/n for the $0$-$5$\% centrality class in $p$/$d$/$^3$He+Au collisions published in \cite{PHENIX:2018lia,Aidala:2018mcw}. The central values were shown previously in our paper \cite{Mace:2018vwq}; the only addition here are error bands with 30\% estimated uncertainties in our computation. These include uncertainties in our choice of $m$, $\mathcal{K}$, $\sigma$,  $T_p^{\rm min}$ as well as possible running coupling corrections; is based in part on the previous detailed study in \cite{Schenke:2013dpa} and the observed sensitivity of our computation to variations in parameters. We have not performed an exhaustive study of these uncertainties; such a study is feasible if time consuming. The biggest systematic uncertainty, as we have indicated, is that of fragmentation. In the absence of a reliable theory framework for fragmentation, this uncertainty is difficult to quantify. In view of this, our estimated uncertainty of 30\% is likely an underestimate. Finally, we should remind the reader of our discussion in Section II of the difference between the measured $v_n\{\rm{EP}\}$ and the computed $v_n\{2\}$; this difference can easily be of the order of 10\%.

Within these caveats, the agreement of the model computation with the PHENIX $v_2$ data is quite good for all three systems in the $0$-$5$\% centrality class. For $v_3$, while the agreement of model with data for $^3$He+Au is very good, the model overshoots the data for both $p$+A and $d$+A collisions.  The analys performed in Ref.~\cite{Mace:2018yvl} based on power counting argument in Eqs.~\eqref{even} and \eqref{odd} makes it clear that 
$v_3$ in contrast to $v_2$ is rather sensitive to variations of  $\mathcal{K}$; indeed, $v_3$  is linearly proportional to  $\mathcal{K}$. Thus small changes of this quantity my improve the agreement between the model results and the data.  
In addition to the stated caveats, since $v_3$ is consistently either above or close to the maximum allowed value of the data, higher order corrections in the $\rho_{\rm pr}/k_\perp^2$ expansion potentially could ``shadow'' $v_3$. In general it will be interesting to understand the impact of such corrections for both $v_2$ and $v_3$ in a simpler model such as the MV model. 

\begin{figure*}[t!]
    \centering
        \begin{subfigure}[t]{0.45\linewidth}
        \centering
	\includegraphics[width=\linewidth]{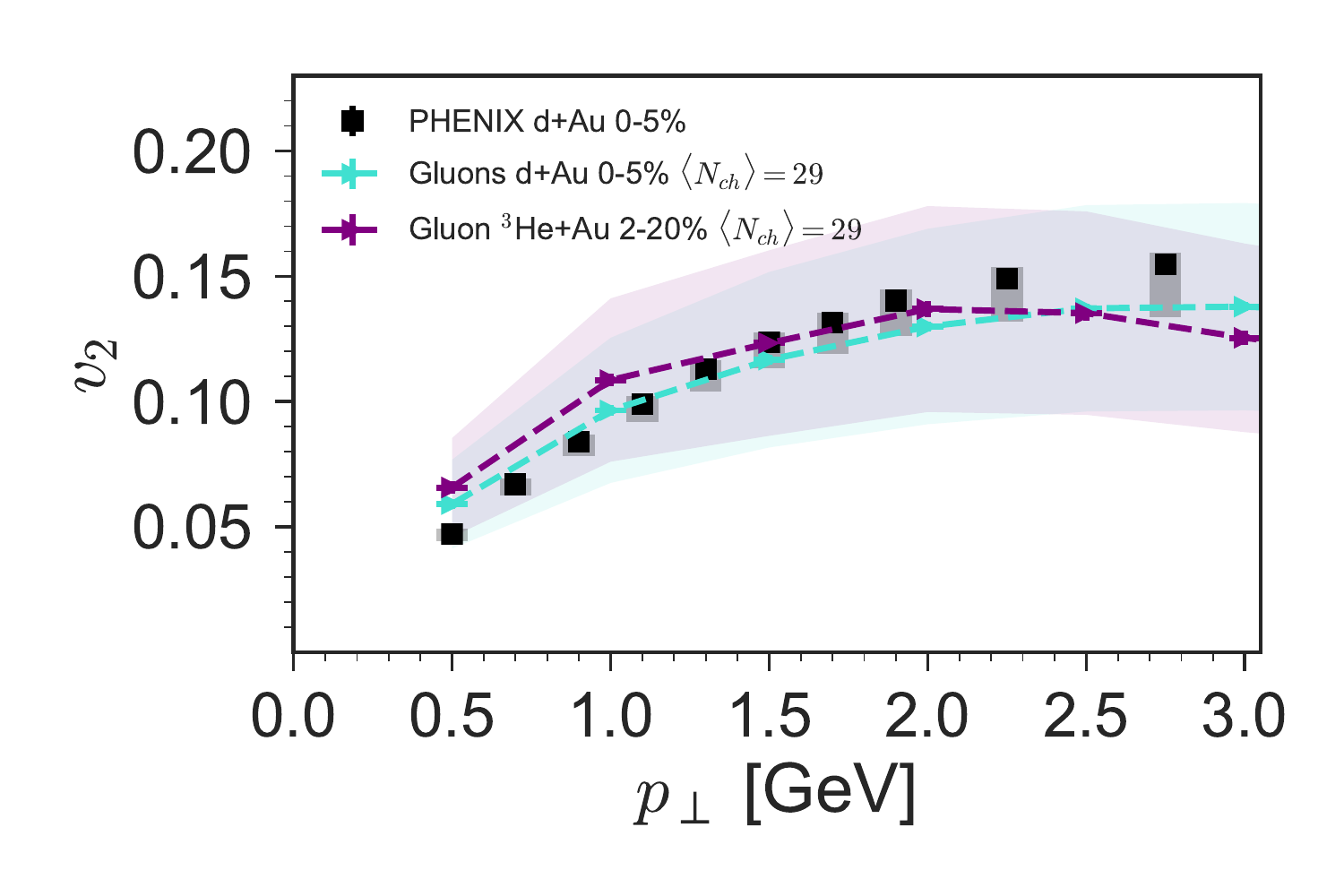}
	\caption{}
\label{fig:same_mult_dhe}
    \end{subfigure}%
    ~
        \begin{subfigure}[t]{0.45\linewidth}
        \centering
	\includegraphics[width=\linewidth]{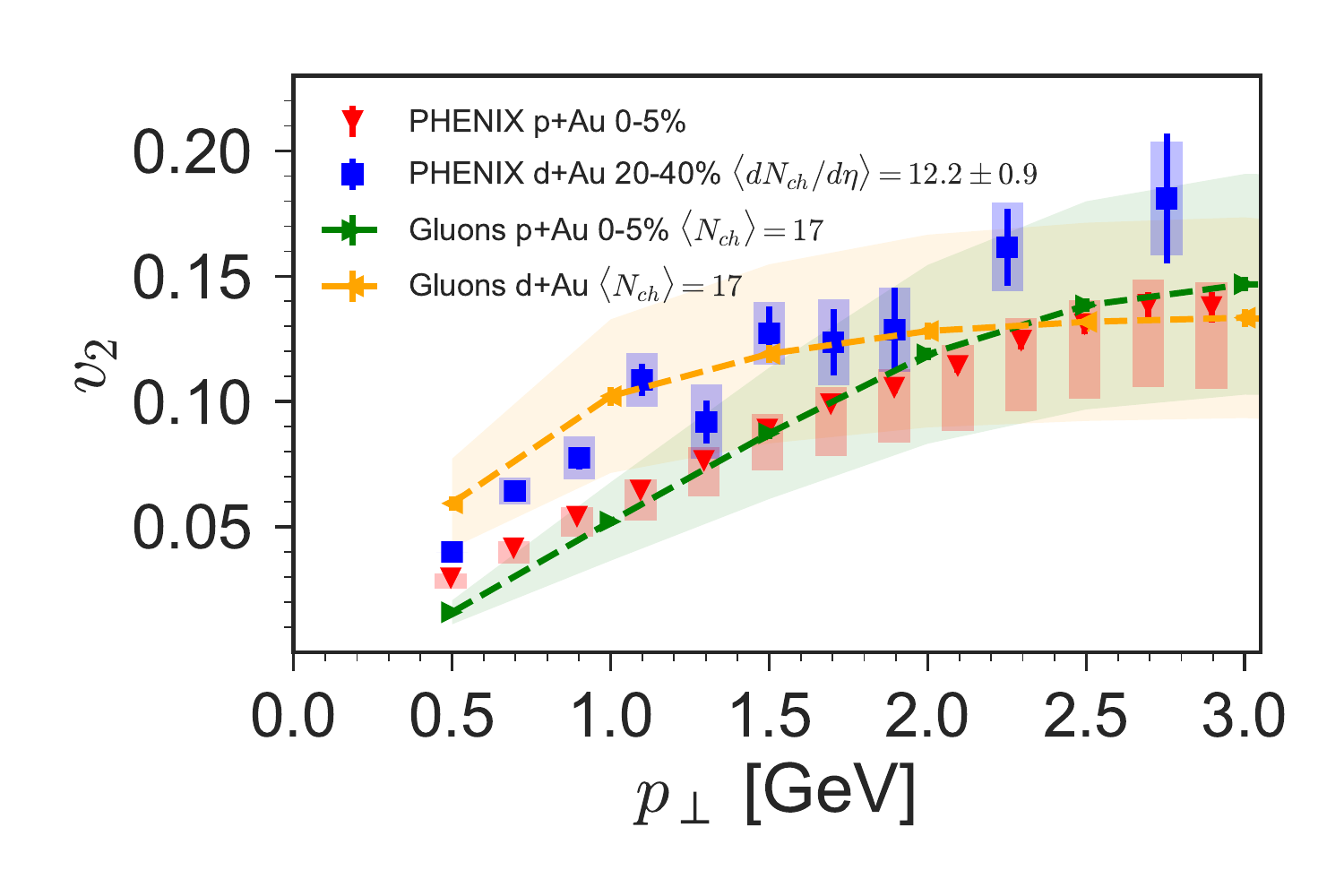}
	\caption{}
	\label{fig:same_mult_pd}
    \end{subfigure}
    \caption{(a) $v_2(p_\perp)$ for d/$^3$He+Au with similar multiplicities for gluons~\cite{Mace:2018vwq}, compared to recent PHENIX results~\cite{PHENIX:2018lia,Aidala:2018mcw} for $d$+Au collisions. Experimental results in the $^3$He+Au centrality class corresponding to $\langle N_{\rm ch} \rangle=29$ are presently not available.  (b) $v_2(p_\perp)$ for $p$/$d$+Au collisions with similar multiplicities for gluons 
    compared to recent PHENIX results~\cite{PHENIX:2018lia,Aidala:2018mcw}. Shaded bands represent the 30\% theory uncertainty. }
\end{figure*}

In \cite{Mace:2018vwq}, we predicted that for the same $N_{\rm ch}$ that $v_{2,3}(p_\perp)$ would be identical for different small systems at fixed $N_{\rm ch}$. This prediction was however based upon a color domain argument rather than explicit computation; as we discussed earlier, the full dilute-dense framework includes HBT and Bose-enhancement contributions which can modify this expectation. In general, we expect the scaling arguments to hold at large $N_{\rm ch}$, which we saw to be the case~\cite{Mace:2018yvl} for $p$+A collisions at the LHC. Fig.~\ref{fig:same_mult_dhe} shows our result for $v_2(p_\perp)$ at a mean $\langle N_g\rangle=29$ corresponding to the mean multiplicity in the $0$-$5$\% centrality class in $d$+Au collisions at RHIC. These results, with the shaded band representing 30\% estimated uncertainty, is compared to the PHENIX data for the same centrality class. Also shown is our prediction for $v_2(p_\perp)$ in $^3$He+Au collisions for the same $\langle N_g\rangle$. We observe that $v_2(p_\perp)$ is nearly identical between the two systems for the same $\langle N_g\rangle$. If the multiplicity distribution for $^3$He+Au from RHIC becomes available, we can map our result for the same in Fig.~\ref{fig:pp_mult} and check whether our prediction is validated by an apples-to-apples data-theory comparison.

In \cite{PHENIX:2018lia,Aidala:2018mcw}, the PHENIX collaboration published data for $v_2(p_\perp)$ for the same $\langle N_{\rm ch}\rangle$ in $p$+Au and $d$+Au collisions corresponding to the $0$-$5$\% centrality class in the former and the $20$-$40$\% centrality class in the latter. Our theory results for the $0$-$5$\% centrality class in $p$+Au collisions (corresponding\footnote{This number, as noted earlier, is consistent with the STAR published data on the multiplicity distribution but larger than the PHENIX value quoted in \cite{PHENIX:2018lia,Aidala:2018mcw}.} to $\langle N_g\rangle=17$) and results for the same $\langle N_g\rangle$ in $d$+Au collisions are compared to the PHENIX data in Fig.~\ref{fig:same_mult_pd}. The bands, as previously, represent 30\% estimated uncertainties. As noted earlier, unlike our results shown in Fig.~\ref{fig:same_mult_dhe}, the results for $p$+Au and $d$+Au for the same $\langle N_g\rangle$ are not identical; this is likely because the $\langle N_g\rangle$ in this latter case are not in the scaling regime. Note that from the inset of Fig.~\ref{fig:all_mults}, the shape of the $p$+Au distribution at high multiplicities is qualitatively steeper than that of $d$+Au/$^3$He+Au. That said, the agreement with the PHENIX data is quite good. It should also be noted that while the theory results for the lower $\langle N_g\rangle$ values clearly differ at low $p_\perp$, the values of $v_2(p_\perp)$ are compatible between the two systems, with the stated theory uncertainty, for $p_\perp > 1$ GeV.

\begin{figure}

\end{figure}

\section{How many-body correlations are generated in colored glass}

The results presented in the previous section suggest that the dilute-dense framework of the CGC presents a competitive description of data in light-heavy ion collisions at RHIC. 
At the very least, they fare no worse than the hydrodynamics based explanations of the data. Unlike hydrodynamic models with MC Glauber-like initial conditions, both the multiplicity distributions  and $v_n$ distributions are generated self-consistently within the model. Further, as we will discuss, the hydro models used in the PHENIX comparisons to data have no clear picture of multiparticle production that can 
be related to the underlying theory. 

In this section, we will delve a little deeper into the underlying systematics of the results presented to arrive at a deeper understanding. Since the n-particle correlations are generated by quantum interference effects, simple intuitive explanations as those available in classical hydrodynamics may be misleading. In \myref\cite{Mace:2018vwq}, we speculated that by comparing the three systems in their respective 0-5\% multiplicity classes, the systems probed different average values of $\Qs$ in the projectile. Larger values of $\Qs$ resulted in larger correlations (at least for $v_2$), which at least qualitatively describes what we saw. However, in reality we are computing solutions to the classical Yang-Mills equations which are highly nonlinear. This is further complicated by more sophisticated modeling of the initial color charge configurations. Thus it is crucial to try to elucidate the salient features of our model which generate the observed correlations. 

In probing the guts of our numerical results, we will examine closely both results from the full model and the MV model wherein analytical and numerical computations are simpler. However it is still difficult to have a complete intuitive understanding of what is happening, because as previously mentioned, quantum interference effects such as  Bose enhancement and Hanbury Brown--Twiss correlations play an important role. We will also take the opportunity here to address misleading interpretations of our work presented in the literature. 

\begin{figure}
\includegraphics[width=0.45\linewidth]{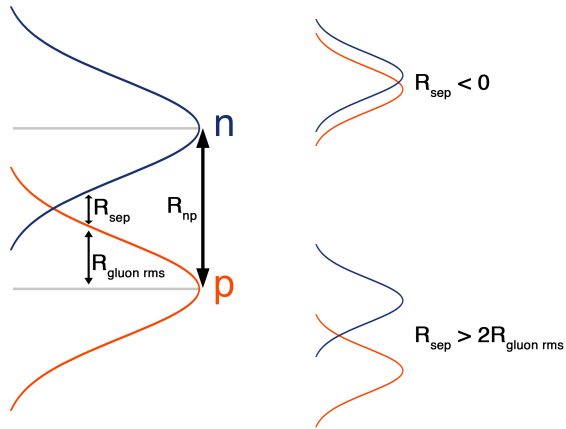}
\caption{Gaussian profiles for the nucleons in the a sampled deuteron configuration, separated center-to-center by distance $R_{\rm np}$. The nucleons have identical widths $R_{\rm gluon-rms}=\sqrt{2B_G}\approxeq 0.56~\text{fm}$.}
\label{fig:deuteron_cartoon}
\end{figure}

For simplicity, we will consider only deuteron-gold collisions in our discussion. Fig.~\ref{fig:deuteron_cartoon} represents the Gaussian profiles for the color charge distributions of the constituent proton and neutron in the deuteron projectile. In the IP-Sat model, the root mean squared gluon radius in the nucleon is $R_{\rm gluon-rms}=\sqrt{2 B_G} = 0.56$ fm~\cite{Caldwell:2010zza}. To understand our results in terms of the distribution of gluon/color charge distributions of projectile interacting with color fields in the gold target, we will define a separation radius $R_{\rm sep}$ which denotes the distance separating the root mean squared widths of the nucleons. When the separation of the peaks of the two nucleons $R_{\rm np}=2 R_{\rm gluon-rms}$, this corresponds to $R_{\rm sep}=0$. Likewise, when the centers overlap completely, $R_{\rm sep}= -2 R_{\rm gluon-rms}$. Thus in this lingo $R_{\rm sep} < 0$ denotes configurations with strong overlap of their nucleon color charge distributions and $R_{\rm sep} >0$ denotes at most that the tails of the gluon distributions overlap.

To understand more deeply the azimuthal anisotropy distributions, it is important to first examine closely how n-body correlations are generated in the first place and how this quantum many-body picture is similar or different from so-called Glauber models\footnote{The Glauber nomenclature is especially ironic since, as we will demonstrate, such models incorrectly treat the Bose statistics of higher order correlations. The proper treatment of such correlations was of course Glauber's fundamental contribution to quantum optics, and as has been noted previously, it is rather the MV model that in this sense embodies the QCD Glauber model.} which discuss such correlations in the language of energy deposition in a collision. As noted, for simplicity, we will examine results in the simpler MV model where the target is approximated by an infinite, uniform nuclear target (no $N_{\rm part}$ fluctuations)~\cite{McLerran:1993ka,McLerran:1993ni}.  
By setting  $\sigma=0$, we will also neglect the fluctuations of the saturation momentum $\Qs$ in  both the projectile and the target.  

In the MV model, simplified analytical calculations were performed in Refs.~\cite{Dumitru:2008wn,Dusling:2009ar,Gelis:2009wh,Kovner:2018azs}. 
In particular, the cumulants of the probability distribution for gluon production were computed recently~\cite{Kovner:2018azs} to leading order in $N_c$ and leading power of terms enhanced by the projectile transverse area. Keeping this leading contribution, the computation in \cite{Kovner:2018azs} showed that the average multiplicity $\kappa_1$ is proportional to   
\begin{equation}\label{fm}
	\kappa_1 \propto 
	S^{\rm pr}_\perp
	 \mu_{\rm pr}^2 
	 \,,
\end{equation}
while the higher order cumulants, which contribute significantly to the high multiplicity tail, are 
\begin{equation}
	\kappa_{n\ge 2} \propto  (n-2)!  
	S^{\rm pr}_\perp
\Lambda^2 
  \left(\frac{ \mu_{\rm pr}^2   }{\Lambda^{2}}\right)^{n}\,.
	\label{Eq:cn}
\end{equation}
Here $\Lambda$ is a nonpeturbative scale of order of the inverse proton radius, $\mu_{\rm pr}^2 \propto Q_{s,{\rm pr}}^2$ is the squared color charge density in the projectile, and $S^{\rm p}_\perp$ is the transverse area of the projectile.  

The factorial growth of the cumulants in Eq.~\eqref{Eq:cn} is specific for the Bose statistics of the gluon degrees of freedom in the underlying theory. 
In Ref.~\cite{Kovner:2018azs}, the authors identified the origin of this factorial growth to be due to Bose enhancement of gluons in the projectile wave function. 
In other phenomenological studies, this factorial growth is often approximated by the {\it ad hoc} postulate of a negative binomial distribution for particle production. In the CGC, we 
can analytically trace the emergence of this approximate negative binomial distribution to the multiparticle Bose statistics of gluons. 

We can study this result further in the context of $d$+Au collisions. If we first consider a typical deuteron configuration in its wave function, the most likely configurations are those with nucleons separated in the transverse plane. For these configurations, we have $S^{\rm pr}_\perp = 2 S^{\rm p}_\perp$ and
$\mu_{\rm pr}^2  =  \mu_{\rm p}^2 $, where   $S^{\rm p}_\perp$ and  $\mu_{\rm p}^2 $ are the transverse area and the squared 
color charge density of a proton, respectively. Thus for ``typical" widely separated nucleon configurations, the cumulants are  
\begin{equation}
	\kappa^{\rm typ}_1 \propto 
	2 S^{\rm p}_\perp
	 \mu_{\rm p}^2 
	\,, 
	\quad 
	 \kappa^{\rm typ}  _{n\ge 2} \propto  2 (n-2)!  
	 S^{\rm p}_\perp
\Lambda^2 
  \left(\frac{ \mu_{\rm p}^2   }{\Lambda^{2}}\right)^{n}\,.
\end{equation}
Now consider rare configurations of the deuteron when the nucleons overlap completely in the transverse plane. 
In this case, we will have $S^{\rm pr}_\perp = S^{\rm p}_\perp$ and $\mu_{\rm pr}^2  =  2 \mu_{\rm p}^2$. 
The corresponding ``rare" cumulants are 
\begin{equation}
	\kappa^{\rm rare}_1 \propto 
	2 S^{\rm p}_\perp
	 \mu_{\rm p}^2 
	 \,, 
	\quad 
	 \kappa^{\rm rare}  _{n\ge 2} \propto  	2^n    (n-2)!  
S^{\rm p}_\perp
\Lambda^2 
  \left(\frac{ \mu_{\rm p}^2   }{\Lambda^{2}}\right)^{n}\,.
\end{equation}

From this simple exercise, we observe that the average multiplicity for both typical and rare configurations 
is the same: $ \kappa^{\rm rare}_1 =  \kappa^{\rm typ}_1$, while the higher order cumulants 
follow a rather different  pattern: $\kappa^{\rm rare}  _{n\ge 2} =   2^{n-1}  \kappa^{\rm typ}  _{n\ge 2}$. In words, the $n$-th order cumulant for $n\geq 2$ 
in rare configurations with overlapping nucleons is $2^{n-1}$  times larger than in typical configurations. This implies that rare overlapping configurations should dominate the high multiplicity tail where higher order correlations contribute significantly.  

\begin{figure}[t]
        \centering
        \includegraphics[width=0.45\linewidth]{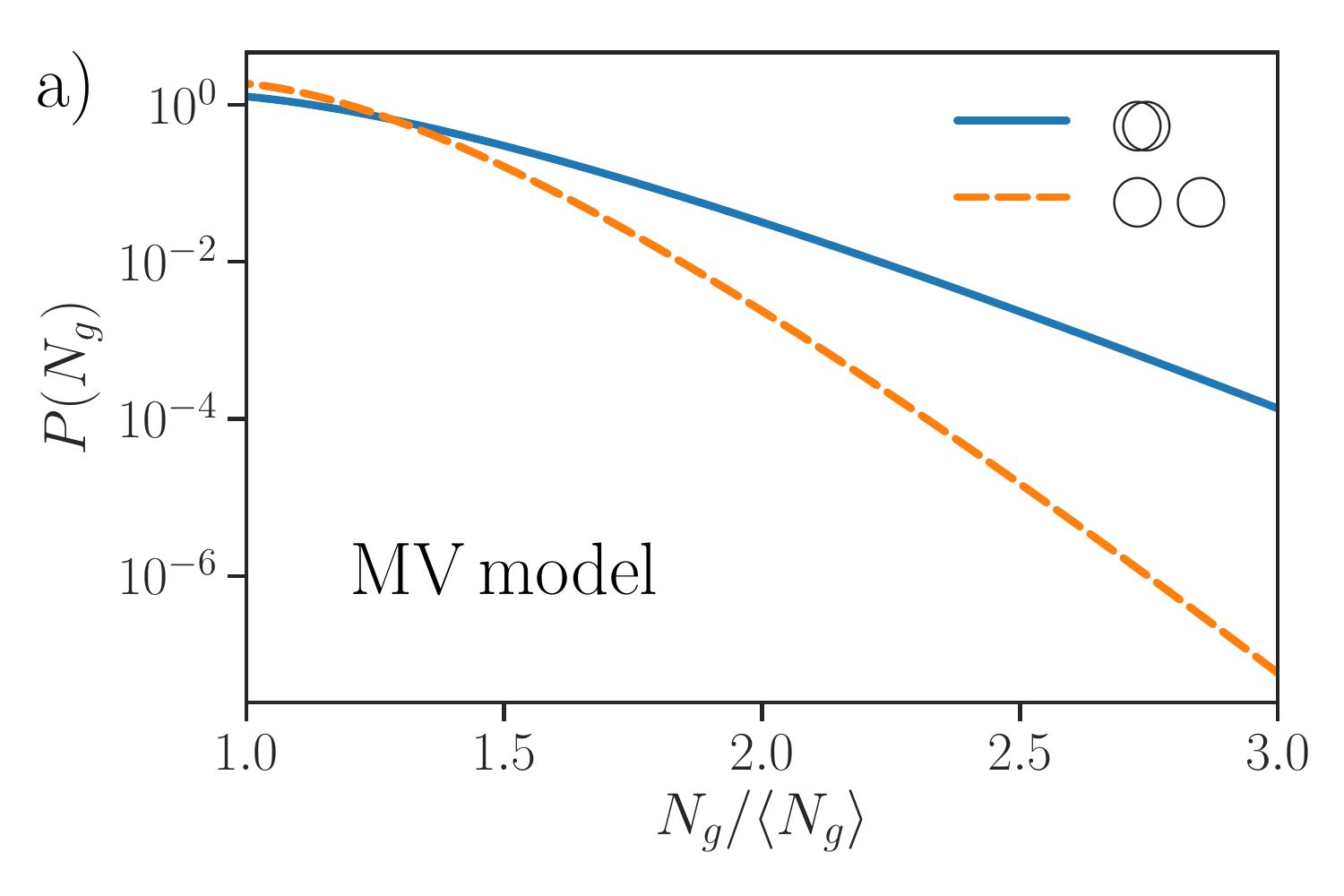}
        \includegraphics[width=0.45\linewidth]{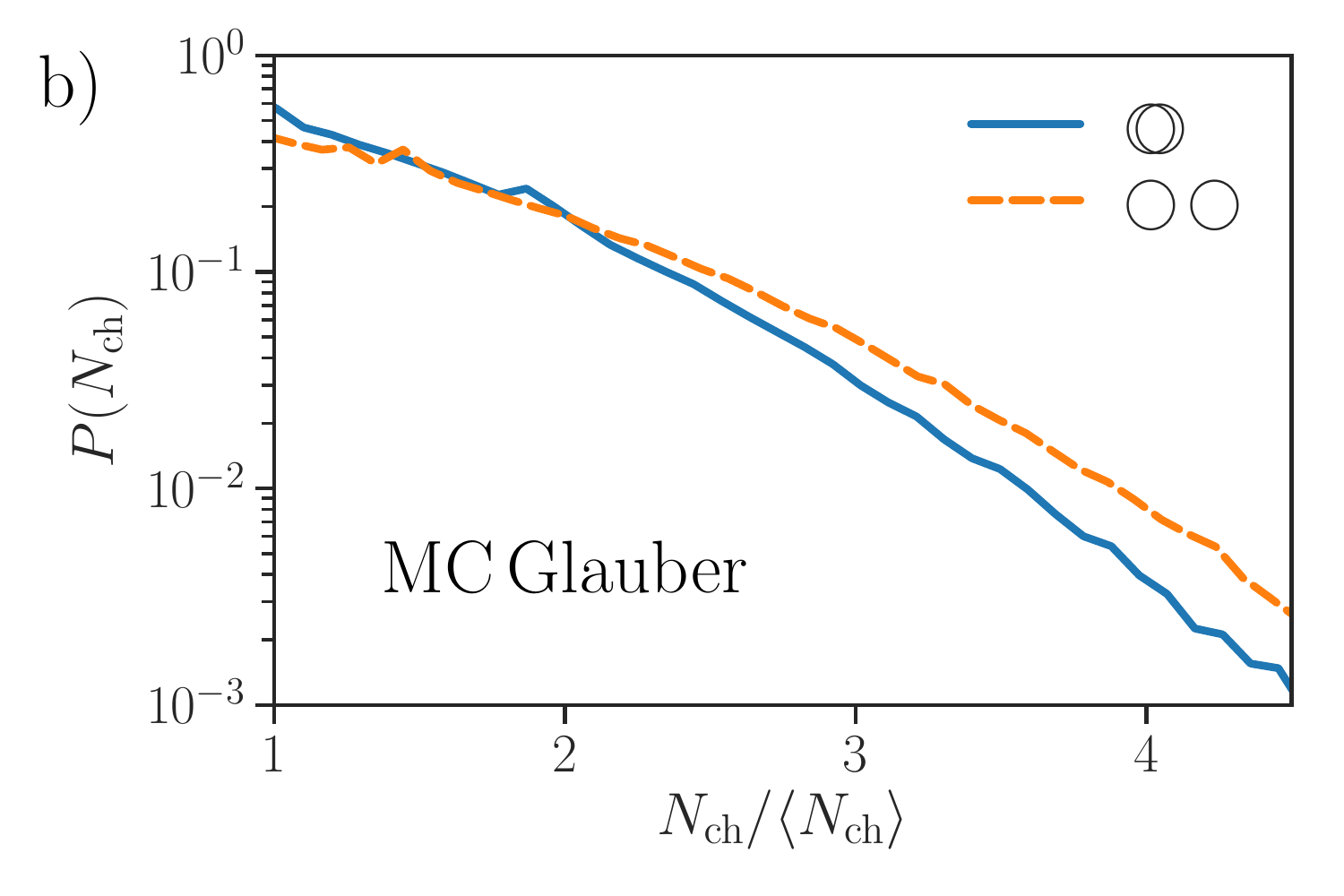}
        \caption{a) The probability $P(N_g)$ in deuteron-MV target collisions for fixed nucleon separation in fully overlapping configurations ($R_{\rm np}=0$) and for configurations where the nucleons are separated in the transverse plane ($R_{\rm np} \gg R_{\rm gluon~radius}$).
	b) The same as a) but in Glauber MC with $x=0.13$ (see App.~\ref{App:MCG} for details).
	}
        \label{fig:prob}
    \end{figure}

To demonstrate this explicitly in the MV model, we computed the n-gluon probability distribution 
for two fixed configurations of the projectile deuteron: fully overlapping and fully separated nucleons in the transverse plane. The results are shown in Fig.~\ref{fig:prob} (a).We observe clearly that closer configurations generate larger multiplicity fluctuations with a fatter probability tail at large $N_g$ thereby fully supporting our simple analytical estimate. 
The enhancement of the high multiplicity tail in the configuration of the deuteron wavefunction with overlapping nucleons is a nontrivial result originating in 
from the Bose enhancement of gluons in QCD that is captured in the CGC EFT. 

This result defies ``classical intuition" expressed by the fact that it is absent in the usual Monte Carlo Glauber  model, the details of which are presented separately in Appendix~\ref{App:MCG}). This MC Glauber model relies on a probabilistic picture of energy distribution but ignores completely the Bose statistics of the color charged gluon constituents.While the lack of fundamental QCD dynamics in the MC Glauber model is not apparent in the average multiplicity, it becomes very visible in the high multiplicity tails that are dominated by the higher order correlations. This is 
seen explicitly in Fig.~\ref{fig:prob} (b). In fact, as clearly seen, the trend in the MC Glauber model for high multiplicity configurations is the opposite of that in the MV model. {\it Ignoring Bose-statistics, or applying it incorrectly, potentially has major consequences in applications of the MC Glauber model to describe any features of higher order correlations in the strong interactions.}

Is the nontrivial quantum nature of gluon correlations observed in the CGC EFT important for the phenomenology of light-heavy ion collisions? One may argue that configurations with overlapping nucleons are suppressed by the available phase space and therefore contribute little to multiparticle production for the specialized case of deuteron-gold collisions. 
This is however not the case as we will now demonstrate. 

\begin{figure}
        \includegraphics[width=0.45\linewidth]{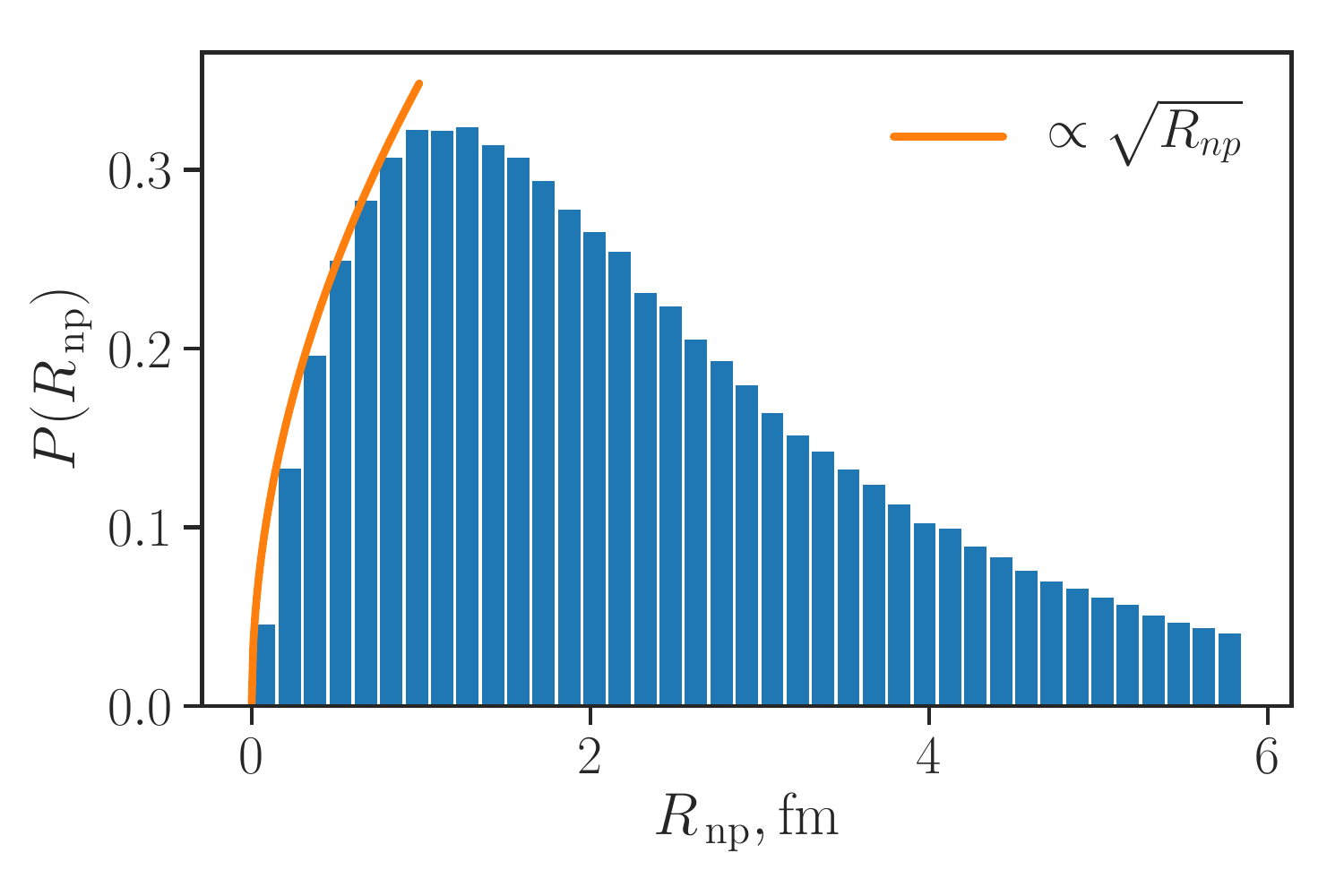}
        \caption{The probability distribution $P(R_{\rm np})$ of transverse separations between the nucleons in the deuteron. The curve added demonstrates that for small $R_{\rm np}$, $P(R_{\rm np})\sim \sqrt{R_{\rm np}}$.}
        \label{fig:glaub}
\end{figure}

Consider the probability distribution describing the transverse separation of nucleons in the deuteron:
\begin{eqnarray}
P(R_{\rm np})= \int d^3x ~\delta(R_{\rm np}-|\v{x}|) P_{\rm np}(\mathbf{x}) \,.
\label{eqn:prob_deut_sep}
\end{eqnarray}
This expression is independent of the target and is only determined by the Hulth\'en deuteron wavefunction~\cite{Miller:2007ri}--indeed,  
\begin{eqnarray}
P_{\rm np}(\mathbf{x})= \phi^2_{\rm np}(|\mathbf{x}|)\,, \qquad {\rm with}\,\, \phi_{\rm np}(r)=\frac{1}{2\pi} \frac{\sqrt{ab(a+b)}}{b-a}\frac{e^{-ar}-e^{-br}}{r}\,,
\label{eqn:prob_hulthen}
\end{eqnarray}
where $a=0.228~\text{fm}^{-1}$ and $b=1.18~\text{fm}^{-1}$~\cite{Adler:2003ii}.

Fig.~\ref{fig:glaub} clearly shows close configurations are suppressed as $P(R_{\rm np})\sim \sqrt{R_{\rm np}}$ at small $R_{\rm np}$. This is however a rather mild suppression relative to the  
exponential enhancement of the CGC high-multiplicity tail in Fig.~\ref{fig:prob}. One may therefore anticipate that the effect of the latter enhancement compensates for the suppression from the former. To study this, we computed the average distance between nucleons at a fixed number of produced gluons in CGC, that is  
\begin{equation}
\label{Eq:RNP}
	\left. \langle R_{\rm np}  \rangle \right|_{N_g}  =   \int dR_{\rm np}\, R_{\rm np}  \, P(R_{\rm np}; N_g) =     \frac{ \int dR_{\rm np}\, R_{\rm np} \, P(R_{\rm np}) \, P(N_g; R_{\rm np})}{  \int dR_{\rm np} \, P(R_{\rm np}) \, P(N_g; R_{\rm np}) }\, .
\end{equation}
Here $P(R_{\rm np}; N_g)$ is the {\it conditional} probability of producing nucleons in the deuteron with separation $R_{\rm np}$ for a fixed number of gluons $N_g$, while $P(N_g; R_{\rm np})$ is the {\it conditional} probability of producing $N_g$ gluons at a fixed distance $R_{\rm np}$ between nucleons in the deuteron. 
In order to get the second equality in Eq.~\eqref{Eq:RNP},  we applied Bayes' theorem. 
Previously, in Fig.~\ref{fig:prob}, we considered two extremes cases for this probability distribution $P(N_g; R_{\rm np}=0)$ and  $P(N_g; R_{\rm np}\gg R_{\rm gluon-rms} )$.  

\begin{figure}[t]
        \centering
\includegraphics[width=0.45\linewidth]{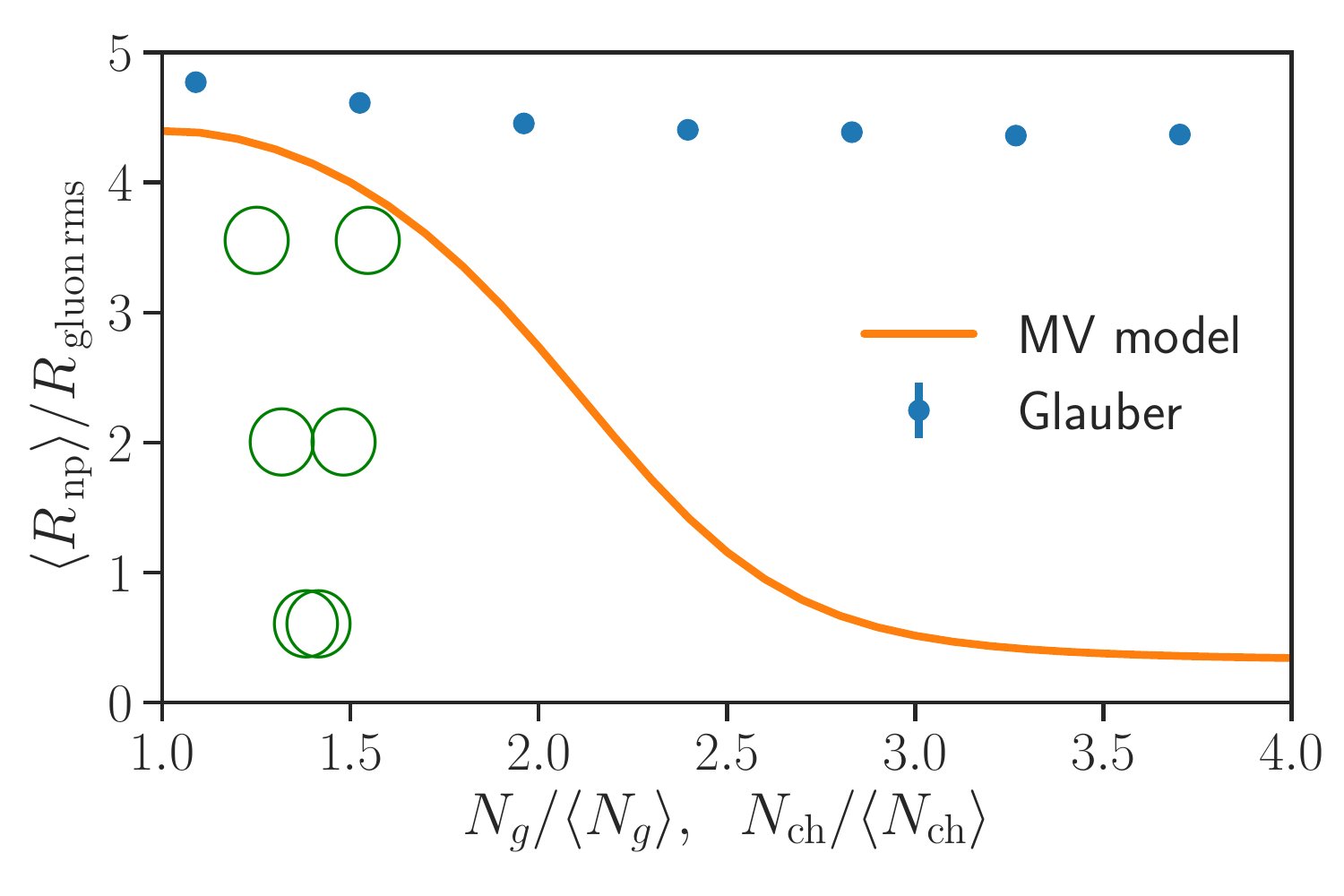}
\caption{Average deuteron transverse separation $\langle R_{\rm np}\rangle$ (normalized by nucleon gluon rms radius $R_{\rm gluon-rms}=0.56$ fm) for $d$-target collisions in the MV model and in the MC Glauber model versus gluon and charged hadron multiplicity. The cartoons denote the color charge overlap of nucleons in the deuteron wavefunction.}
\label{fig:averd}
    \end{figure}

In Fig.~\ref{fig:averd}, we show the multiplicity biased average $\langle R_{\rm np}  \rangle$  as a function of $N_g$, the number of produced gluons. (Note that $R_{\rm sep}$ that we defined 
previously satisfies $R_{\rm sep}= R_{\rm np} - 2 R_{\rm gluon-rms}$.) The figure demonstrates clearly the anticipated dependence of this ratio on the number of produced gluons $N_g$ normalized by the average gluon multiplicity $\langle N_g\rangle$. What is most striking however is that the behavior of this ratio is quite different from the corresponding calculation in the MC Glauber model; for the latter, the multiplicity bias of low and high multiplicities relative to the average is virtually independent of the distance between nucleons. 

We emphasize that the average {\it unbiased} distance between nucleons is the same in the MV model calculations and in the MC Glauber model because it depends only on the parameters of the  Hulth\'en deuteron wavefunction.  In summary, albeit configurations with overlapping nucleons are suppressed by the available phase space,  Bose statistics enhanced higher order correlations present in the CGC lead to a significant contribution of these configurations (in contrast to the MC Glauber model) in the high multiplicity tails of $d$+Au collisions\footnote{Note that in the IP-Glasma+MUSIC hydrodynamic model, we expect the same spatial distributions in high multiplicity tails as seen in the dilute-dense framework.}. 

\begin{figure}
\includegraphics[width=0.48\linewidth]{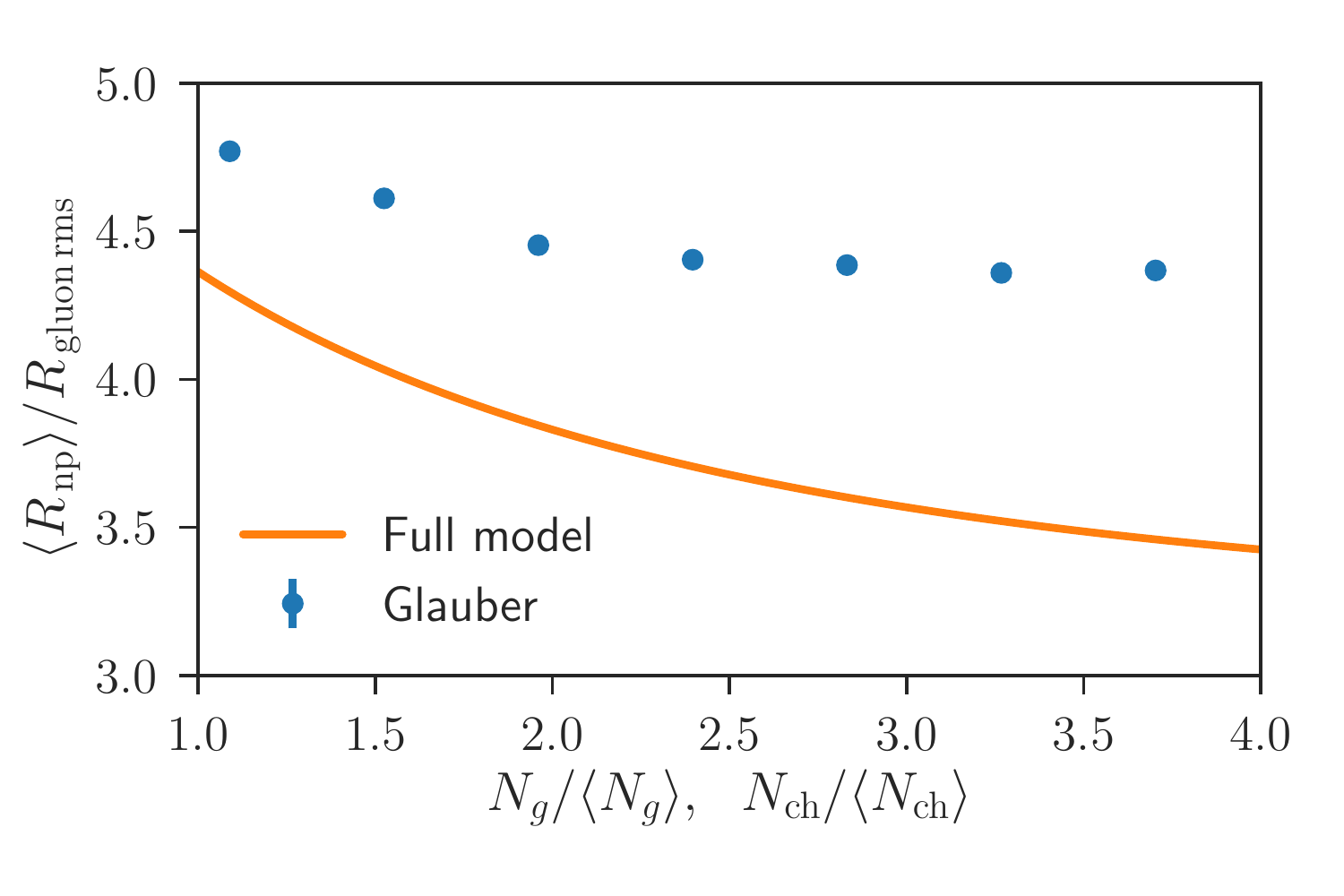}
\caption{Average deuteron transverse separation $\langle R_{\rm np}\rangle$ (normalized by nucleon gluon rms radius $R_{\rm gluon-rms}$) for $d$+Au collisions in the event-by-event dilute-dense CGC model and in the MC Glauber model versus gluon and charged hadron multiplicity.  }
\label{fig:dvsN_dilutedense}
\end{figure}

Restoring target fluctuations and $\Qs$ fluctuations, the corresponding full model shows a similar dependence on $R_{\rm sep}$ and $N_{\rm ch}$. However in this case, the systematics observed in Fig.~\ref{fig:averd} must be convoluted with the probability that the nucleons in the deuteron collide with nucleons the target nucleus. This weakens the anti-correlation observed between nucleon separation and gluon multiplicity. Our results are shown in Fig.~\ref{fig:dvsN_dilutedense}; a clear suppression of $\langle R_{\rm np}\rangle$ in the dilute-dense CGC relative to MC Glauber is seen with increasing multiplicity.

We have emphasized the quantum statistics of gluons as the dominant cause of the  nontrivial many-body correlations that enhance contributions to the tails of multiplicity distributions 
from overlapping nucleon configurations. Such quantum correlations can be diluted by increasing  the number of the species participating in the quantum interference effects. 
For gluons, the easiest way to achieve this is by taking the limit of large number of gluon color charge species -- corresponding to $N_c\rightarrow \infty$.  An example of this is the correlated part of the double inclusive production cross-section; in the CGC EFT, it originates from a nontrivial cross-talk between the projectile/target color sources in both factors on the r.h.s of Eq.~\eqref{eq:Glitter}. This is a manifestation of the fact that the connected part of two gluon production is suppressed by a factor of $N_c^2-1$ compared to the square of the single inclusive gluon production. 

Indeed, it is critical to observe that if the correlation between the factors in Eq.~\eqref{eq:Glitter} is neglected (in other words, if one takes the  strict infinite $N_c$ limit), one will obtain 
\begin{align}
\lim_{N_c\to \infty } \frac{d^{2}N}{d^{2}k_{\perp,1}dy_{1}d^{2}k_{\perp,2}dy_{2}} &= 
\lim_{N_c\to \infty }
\left\langle \left \langle
\frac{dN}{d^{2}k_{\perp,1}dy_{1}}  \Big[\rho_{\rm pr}, \rho_{\rm t}\Big]
\frac{dN}{d^{2}k_{\perp,2}dy_{2}}  \Big[\rho_{\rm pr}, \rho_{\rm t}\Big]
 \right\rangle_{\rm pr} 
 \right\rangle_{\rm t} \\ 
 &= 
 \left\langle \left \langle
\frac{dN}{d^{2}k_{\perp,1}dy_{1}}  \Big[\rho_{\rm pr}, \rho_{\rm t}\Big]
 \right\rangle_{\rm pr} 
 \right\rangle_{\rm t} 
  \left\langle \left \langle
\frac{dN}{d^{2}k_{\perp,2}dy_{2}}  \Big[\rho_{\rm pr}, \rho_{\rm t}\Big]
 \right\rangle_{\rm pr} 
 \right\rangle_{\rm t} =  
 \frac{d N}{d^{2}k_{\perp,1}dy_{1}}
 \frac{d N}{d^{2}k_{\perp,2}dy_{2}} \notag\,. 
\end{align}
Two-particle correlations and associated harmonics are strictly zero in this limit. Higher order cumulants will likewise vanish impacting the physics of Bose correlations in the high multiplicity tails. 

In a recent work~\cite{Nagle:2018ybc}, the authors used a formula for the average multiplicity in dilute-dense collisions  that asymptotes\footnote{In the dilute-dense framework, 
the average multiplicity asymptotes to $\langle N_g\rangle = Q_{s,\rm{pr}}^2 \ln(Q_{s,t}^2/Q_{s.{\rm pr}}^2)$; further it is not clear that the $m$ in \cite{Nagle:2018ybc} is identical to the cutoff scale 
$m$ that regulates the Poisson equation in this work.} to the form $\langle N_g\rangle = Q_{s,\rm{pr}}^2 \ln(Q_{s,t}^2/m^2)$ where they choose m=0.3 GeV. By varying $Q_{s,{\rm pr}}$ and $Q_{s,t}$ from impact parameter fluctuations {\it a la} MC Glauber, as well as additional $Q_s$ fluctuations to multiplicity distributions, they extracted quite different values of $Q_s$ from those discussed here. However color charge fluctuations are not included in \cite{Nagle:2018ybc}; this is therefore equivalent to working in the $N_c=\infty$ limit discussed above where $\langle N_g N_g\rangle = \langle N_g \rangle \, \langle N_g\rangle$. Any n-particle correlations that arise come only from event-by-event fluctuations in $Q_s$ from impact parameter fluctuations and the additional $Q_s$ fluctuations included in \cite{Nagle:2018ybc} (with the physics interpretation of the latter now unclear). 

Thus effectively even though a ``CGC formula" for the average multiplicity is employed in \cite{Nagle:2018ybc}, all higher order correlations are MC Glauber model-like. From our previous discussion, it is clear that this model misses the Bose enhancement of $n \geq 2$ cumulants in high multiplicity tails (the very contributions that Glauber showed distinguishes quantum optics from its classical counterpart) leading to a qualitatively different picture of light-heavy ion collisions at ultrarelativistic energies. Further, there are no intrinsic correlations that are long range in rapidity in this picture like those between the color sources in the projectile and target that exchange color over a wide 
rapidity range. Any such correlations, if they exist, are put in by hand, leaving their dynamical origins obscure. Long range rapidity correlations  generated by final state interactions are limited to the horizon provided by the sound speed~\cite{Staig:2010pn,Springer:2012iz}.

\begin{figure}
\includegraphics[width=0.75\linewidth]{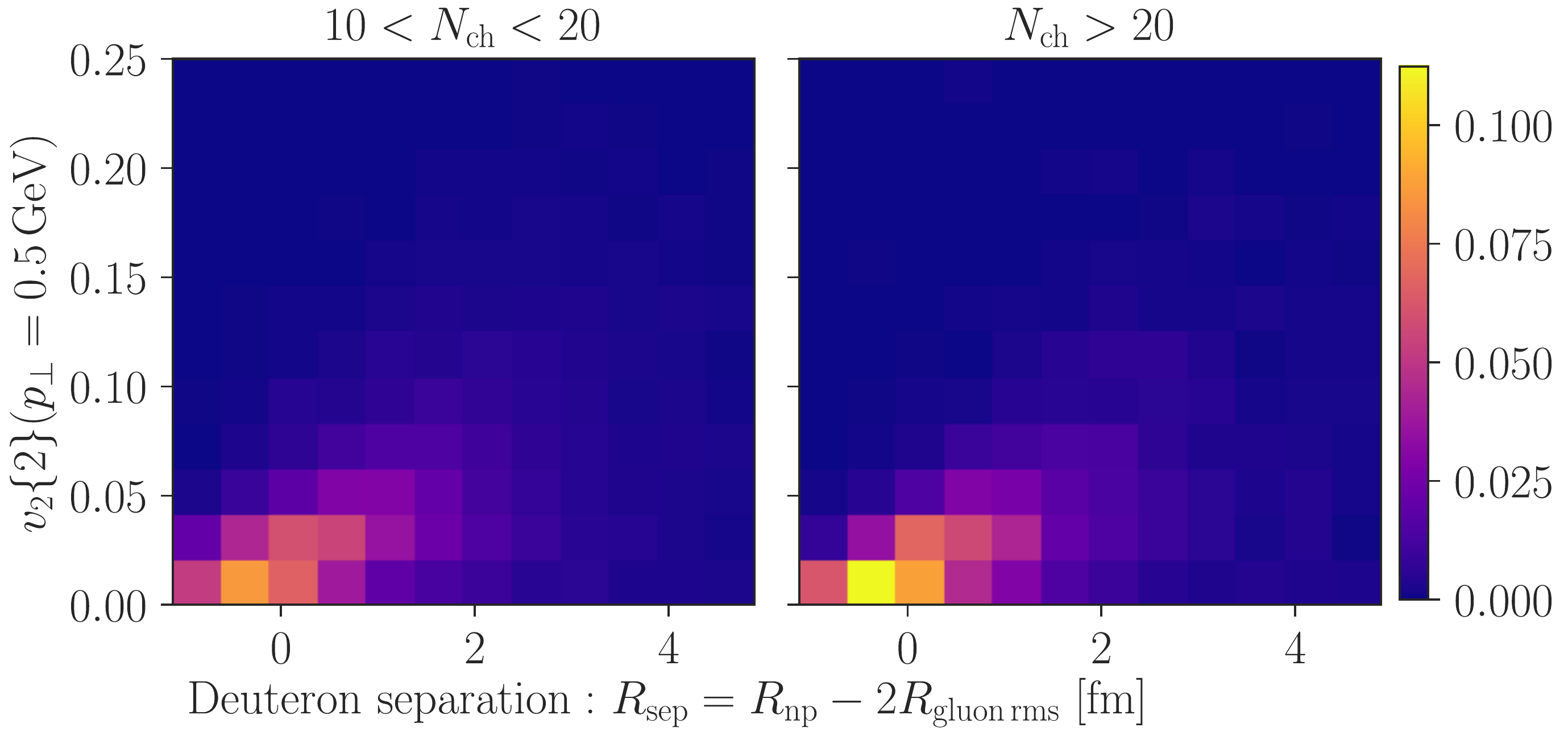}
\caption{Histogram of $v_2\{2\}(p_\perp=0.5~\text{GeV})$ as a function of the deuteron separation. Bin colors denote the probability of obtaining the shown value of $v_2$ at a given separation. }
\label{fig:d_hist_05}
\end{figure}

\begin{figure}
\includegraphics[width=0.75\linewidth]{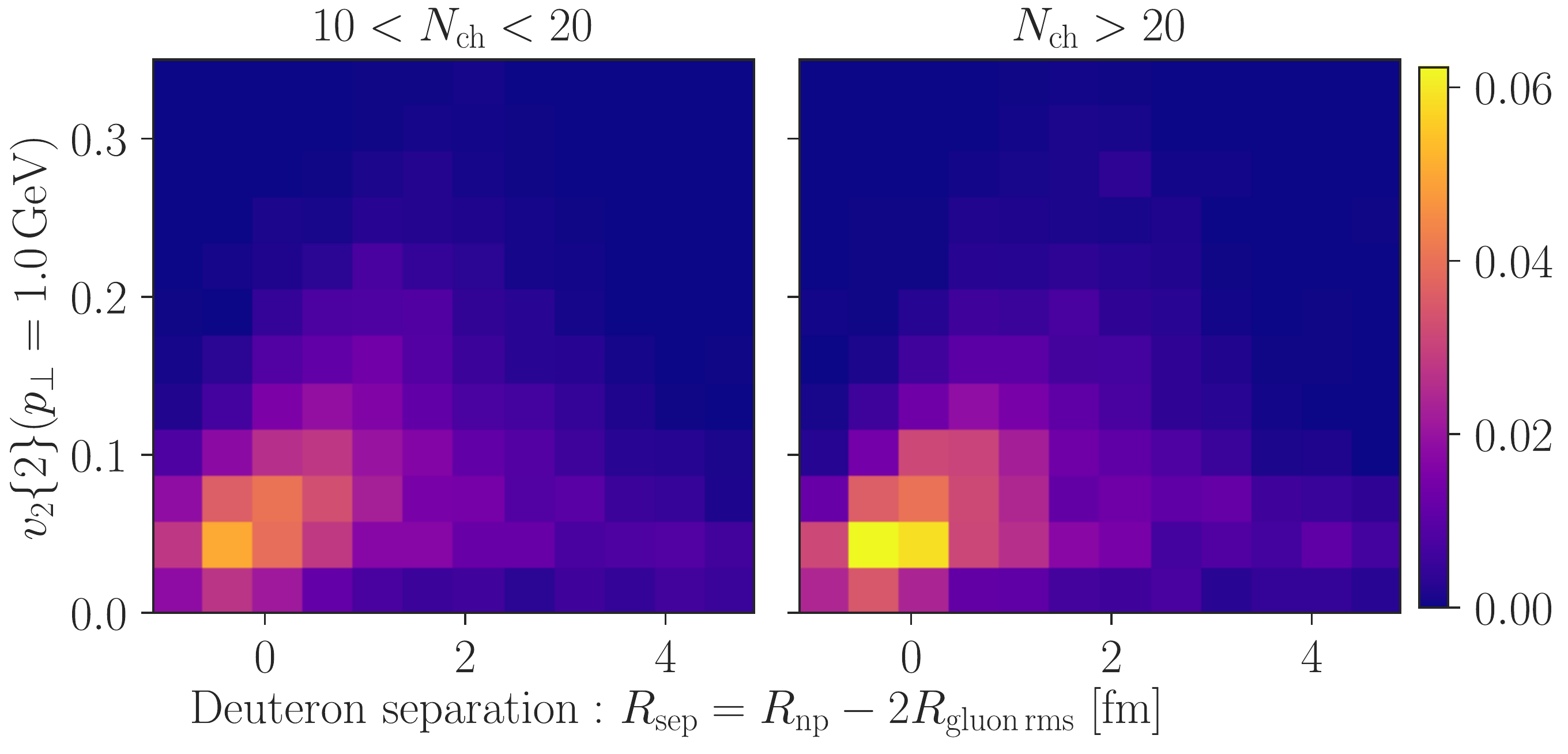}
\caption{Similar to Fig.~\ref{fig:d_hist_05}, but for $p_\perp=1~\text{GeV}$.}
\label{fig:d_hist_1}
\end{figure}

\begin{figure}
\includegraphics[width=0.75\linewidth]{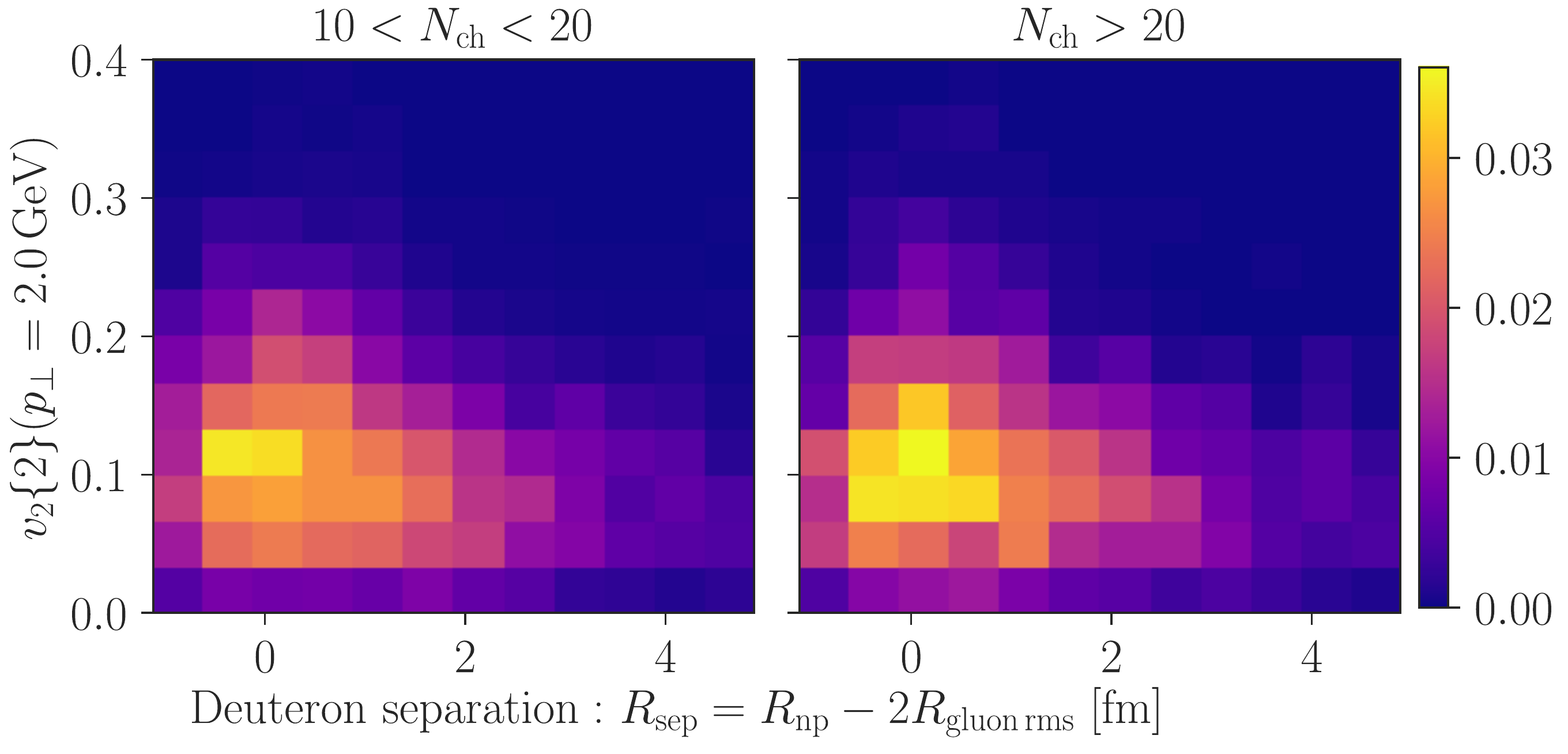}
\caption{Similar to Fig.~\ref{fig:d_hist_05}, but for $p_\perp=2~\text{GeV}$.}
\label{fig:d_hist_2}
\end{figure}

We turn now to a more detailed examination of the $v_2$ azimuthal anisotropy in our dilute-dense CGC framework. 
We will examine both high and low multiplicity events and label events according to how the nucleon positions are distributed in $R_{\rm sep}$. For all events with a given $N_g$ and $R_{\rm sep}$, we will plot the event-by-event distribution of $v_2$ for a given $p_\perp$. The resulting histograms of these (for $10 < N_g < 20$ and  $N_g > 20$) with $v_2$ computed at $p_\perp = 0.5$, $1$ and $2$ GeV are shown in Figs.~\ref{fig:d_hist_05}-\ref{fig:d_hist_2}. Note that the choices of multiplicity bin do not directly correspond to percentile multiplicity bins, and that the scale for the probability changes between the different momenta considered.

\begin{figure}
\includegraphics[width=0.47\linewidth]{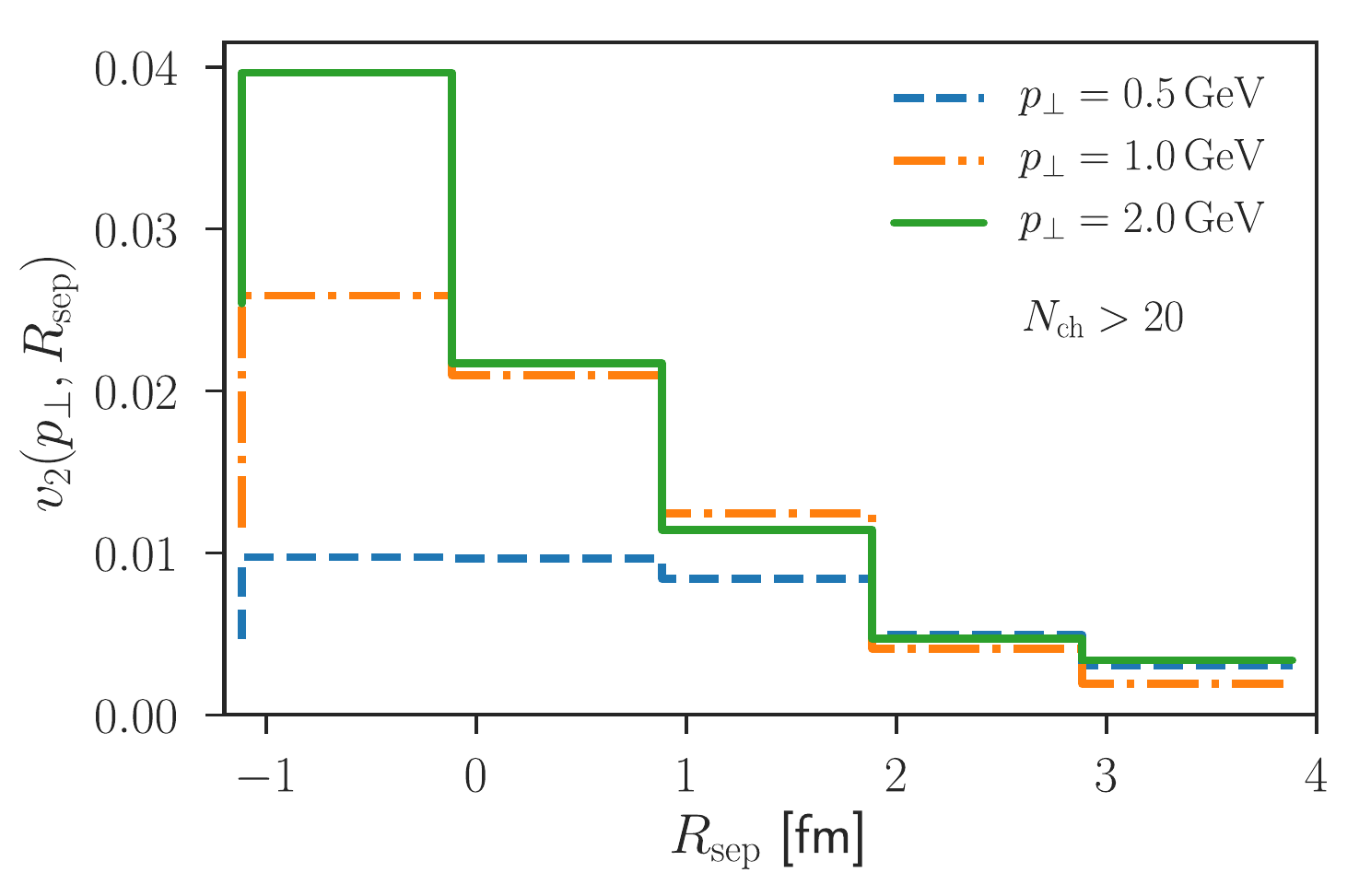}
\caption{Plot of $v_2(p_\perp,R_{\rm sep})$ versus $R_{\rm sep}$ for different values of $p_\perp$.}
\label{fig:v2-Rsep}
\end{figure}

For $v_2(p_\perp=0.5\, {\rm GeV})$, we see that there are significant contributions to $v_2$ from overlapping configurations. For the larger $p_\perp$ values one sees contributions from a wide distribution $R_{\rm sep}$. The histograms themselves may be misleading because the $v_2$'s shown should be convoluted with the probability of having separations of a given $R_{\rm sep}$. 
In Fig.~\ref{fig:v2-Rsep}, we plot 
\begin{equation}
v_2(p_\perp,R_{\rm low} < R_{\rm sep} < R_{\rm high}) = \sum_{v_2(p_\perp)} \sum_{{R}_{\rm sep}=R_{\rm low}}^{R_{\rm high}} v_2(p_\perp) P(v_2(p_\perp),{R}_{\rm sep}) \,,
 \end{equation}
versus bins in $R_{\rm sep}$ for $p_\perp=0.5$, $1.0$ and $2.0$ GeV. On the r.h.s of this expression, $P(v_2(p_\perp),{R}_{\rm sep})$ denotes the probability for a given value of the azimuthal anisotropy $v_2(p_\perp)$ for configurations with separations in a range $R_{\rm low} < R_{\rm sep} < R_{\rm high}$. The sum of the expression over all $R_{\rm sep}$ corresponds to our theory result for $v_2(p_\perp)$. All of this information is contained in principle in the histograms in Figs.~\ref{fig:d_hist_05}-\ref{fig:d_hist_2}; however Fig.~\ref{fig:v2-Rsep} makes clear the fact that $v_2$ shown in the histograms should be appropriately weighted to to draw any conclusions.

Fig.~\ref{fig:v2-Rsep} shows a nontrivial pattern. The largest contribution to the azimuthal anisotropy coefficient $v_2$ at all the momenta considered come from closely overlapping color charge configurations of the nucleons in the projectile. This is striking because, while as Fig.~\ref{fig:dvsN_dilutedense} suggests such configurations are more likely in rare events, they are nevertheless not the dominant configurations. Further, this is contrary to what one would in principle expect from hydrodynamics where large momentum anisotropies result from large spatial eccentricities corresponding to configurations with the nucleons widely separated. Our results therefore clearly indicate an alternative mechanism for generating $v_2$. The effect in overlapping configurations is more pronounced at the higher $p_\perp$ values of $2$ GeV. As shown in \cite{Dumitru:2010iy,Dusling:2013qoz}, and discussed at length in \cite{Dusling:2012wy}, near and far side azimuthal correlations are generated when the unintegrated gluon distributions in the projectile and target overlap maximally in transverse momenta; this can also be seen from the structure of Eqs.~\eqref{eqn:dNevenodd}, \eqref{eq:vn} and \eqref{eq:vn-formula}. Since the typical transverse momenta in projectile and target are respectively of the order of the corresponding $Q_s$, for rare configurations one expects the azimuthal collimation to be maximal for momenta of the order of a few GeV. In future work, we  plan to look more closely in the MV model into the systematics of this effect in light-heavy collisions.

\section{Summary and Outlook}
In this paper, we discussed the computation of n-gluon correlations in the dilute-dense CGC framework. Our work significantly extends the discussion of the $v_n$ azimuthal anisotropy coefficients in hadron-nucleus collisions outlined in \cite{Mace:2018vwq,Mace:2018yvl}. We focused here on light-heavy ion collisions at the highest RHIC energies and showed that our framework provides a 
competitive description of the $v_{2,3}(p_\perp)$ data on $p$/$d$/$^3$He+Au collisions published by the PHENIX collaboration~\cite{PHENIX:2018lia,Aidala:2018mcw}. We also showed here that our framework provides a good description of the $v_2$ data published by PHENIX comparing $v_2(p_\perp)$ in $p$+Au collisions in the $0$-$5$\% centrality class to $v_2(p_\perp)$ in $d$+Au collisions in the 
$20$-$40$\% centrality class that corresponds to the same $\langle N_{\rm ch}\rangle$. We also predicted the behavior of $v_2(p_\perp)$ in $d$+Au and $^3$He+Au collisions for the same $\langle N_g\rangle$. This result however assumes that the multiplicity distribution in $^3$He+Au collisions can be described with the same choice of free parameters that we employed in fitting the 
$d$+Au multiplicity distribution. 

The essential feature of our dilute-dense CGC framework is the event-by-event solution of the classical QCD Yang-Mills equations for a given distribution of light-cone color charge densities in both the projectile and nuclei. The procedure followed, and outlined here in some detail, is similar to that followed in the dense-dense CGC framework exemplified by the IP-Glasma model. As we discussed, the primary difference is that in the dilute-dense framework the color charge density is expanded to lowest nontrivial order in the projectile while retained to all orders in the target; no such approximation is made in the dense-dense framework. The dilute-dense approximation provides analytical expressions that are simpler to compute. For instance, unlike the dense-dense case, it is not necessary to solve the Yang-Mills equations in the forward light-cone; this makes numerical computations in the dense-dense framework challenging. A task for future work is to estimate the effect of higher order corrections in the color charge density of the projectile. Another interesting project would be to compute azimuthal anisotropy coefficients from four and higher particle correlations. 

Even though the dilute-dense framework is simpler, one is still solving nonlinear equations and computing quantum interference contributions to n-gluons correlations. Simple analytical insight into the push and pull of different effects is therefore not readily available. We nevertheless looked closely into the structure of n-gluon correlations and showed that the high multiplicity tail in 
multiparticle production is dominated by Bose enhanced higher order cumulants. A consequence, which we discussed in the specific case of deuteron-gold collisions, is that high multiplicity tails are dominated by rare overlapping configurations of nucleons while more typical low multiplicity events correspond to nucleon configurations where the nucleons are separated. This is in qualitative contrast to the MC Glauber model where configurations where nucleons are separate dominate both low and high multiplicity events. We argued that is is because the MC Glauber model does not include Bose enhancement effects that have to be included in any theory with boson degrees of freedom and points to a fundamental limitation in employing this framework to describe higher order correlations in such theories. 

There are still a number of features of the systematics of azimuthal anisotropies in the dilute-dense framework that need to be better understood. One can also not minimize the uncertainties that arise in comparison of this theory framework to data. As we emphasized previously, the most significant among these is our lack of understanding of how gluons fragment to hadrons. 
This is a fundamental problem in QCD and the various approaches from string fragmentation to instantaneous thermal freeze-out are prescriptions that are at best constrained by fits to data. However the applicability of such prescriptions to rare events in small systems is unclear. Another uncertainty in our treatment is the role of final state rescattering in modifying correlations that are present in the initial state. With increasing multiplicity, final state rescattering cannot be neglected, especially with increasing system size. How this occurs is clearly a problem that deserves better understanding. Finally, while the IP-Sat model has the virtue of being constrained by fits to HERA inclusive and exclusive data, the JIMWLK framework (and refinements thereof) is the appropriate one at small x; implementation of this framework however requires a better understanding of the evolution of spatial color charge distributions and their moments with decreasing x~\cite{Mantysaari:2018zdd}.

\acknowledgements
This work is dedicated to the memory of Prof. Roy Glauber whose pioneering work emphasized the importance of photon statistics in the quantum theory of optical coherence. 
We thank Kevin Dusling, Alex Kovner, Yuri Kovchegov, Jiangyong Jia, Tuomas Lappi, Wei Li, Larry McLerran, Bjoern Schenke, and Chun Shen for stimulating discussions. We thank Bjoern Schenke for very helpful comments on the manuscript.
We would like to acknowledge the support, in part, by the U.S. Department of Energy, Office of Science, Office of Nuclear Physics, under Contracts No.
DE-SC0012704 (M.M.,P.T.,R.V.) and DE-FG02-88ER40388 (M.M.). M.M. and V.S.  would also like to thank the BEST DOE Topical Theory Collaboration for support. The work of M.M. was supported in part by
the European Research Council, grant ERC-2015-CoG681707.  This research used resources
of the National Energy Research Scientific Computing Center, a DOE Office of Science User Facility supported by the Office of Science of the U.S. Department
of Energy under Contract No. DE-AC02-05CH11231. 

\appendix

\section{MC-Glauber} 
\label{App:MCG}
For completeness we describe the  MC-Glauber model we used to compare it with the MV model calculations.  
Our implementation of the MC-Glauber simulation based on a two-component model~\cite{Kharzeev:2000ph} can be described as follows. At a given impact parameter, we assume a collision to generate $x N_{\rm coll} + (1-x) \frac{N_{\rm part}}{2}$ number of identical sources, each of which produces ``n" particles following a negative-binomial distribution of mean $\bar{n}$ and width $\sim 1/k$:
\begin{equation}\label{eq:nbd}
  P_n^{\rm NBD}(\bar{n},k) = \frac{\Gamma(k+n)}{\Gamma(k)\Gamma(n+1)} \frac{\bar{n}^n k^k}{(\bar{n}+k)^{n+k}}\,.
\end{equation}

The value of $\bar n=2.05$ at $\sqrt{s_{NN}}=200$ GeV is obtained from the parameterization of charged multiplicity in nonsingle diffractive $\bar{p}p$ interactions provided in Ref.~\cite{Abe:1989td}. Here, $N_{\rm coll}$, $N_{\rm part}$ and $x$ are the number of binary collisions, the number of participating nucleons and the ``hardness'' scale, respectively. Both $x=0.13$ and $k=2$ are free parameters that are tuned by fitting the min-bias charged hadron multiplicity distribution in $d$+Au collisions~\cite{Abelev:2008ab}. The final input to our MC-Glauber model is the inelastic $p$+$p$ collision cross section of $\sigma_{NN}^{inel}=42$ mb at $\sqrt{s_{NN}}=200$ GeV that is constrained by the global data on total~\cite{Agashe:2014kda} and elastic~\cite{Antchev:2011vs} $p$+$p$ collisions.

\bibliography{bibl.bib}

\end{document}